\documentclass[reprint,pre,aps,twocolumn,amsmath,amssymb]{revtex4-2}

\usepackage{graphicx}
\usepackage{bm,subfigure,color,url,booktabs,mathrsfs}%cite
\usepackage[colorlinks=true,citecolor=myblue,linkcolor=myblue,urlcolor=myblue]{hyperref}
\usepackage{float}
\restylefloat{table}

\newcommand{\bhline}[1]{\noalign{\hrule height #1}}

\definecolor{myblue}{RGB}{0,0,200}

\begin{document}

\title{
Conjugate Distribution Laws in Cultural Evolution via Statistical Learning
}

\author{Eita Nakamura}
\email{eita.nakamura@i.kyoto-u.ac.jp}
\affiliation{The Hakubi Center for Advanced Research, Kyoto University, Sakyo, Kyoto 606-8501, Japan}
\affiliation{Graduate School of Informatics, Kyoto University, Sakyo, Kyoto 606-8501, Japan}

\date{September 8, 2021}
%\date{\today}

\begin{abstract}
Many cultural traits characterizing intelligent behaviors are now thought to be transmitted through statistical learning, motivating us to study its effects on cultural evolution. We conduct a large-scale music data analysis and observe that various statistical parameters of musical products approximately follow the beta distribution and other conjugate distributions. We construct a simple model of cultural evolution incorporating statistical learning and analytically show that conjugate distributions emerge at equilibrium in the presence of oblique transmission. The results demonstrate that the distribution of a cultural trait within a population depends on the individual's model for cultural production (the conjugate distribution law), and reveal interesting possibilities for theoretical and experimental studies on cultural evolution and social learning.
\end{abstract}

\maketitle

%%%%%%%%%%%%%%%%%%%%%%%%%%%%%%%%%%%%%%%%%%%%%%%%%%%%%%%
\section{Introduction}
%%%%%%%%%%%%%%%%%%%%%%%%%%%%%%%%%%%%%%%%%%%%%%%%%%%%%%%

Cultural transmission and evolution are essential to the development of human society \cite{Szathmary1995}.
The common approach to studying these processes is to analyze dynamical systems or stochastic processes that incorporate the transmission, selection, and mutation processes of cultural traits \cite{CavalliSforzaFeldman,BoydRicherson}.
The well-developed theories of genetic evolution and population dynamics can be applied for understanding the dynamics of some cultural traits such as an individual's native language (English/French/etc.) \cite{CavalliSforzaFeldman,Nowak1999,Abrams2011}.
However, many cultural traits characterizing intelligent behaviors such as speech (accent, speech speed, etc.) are transmitted in a manner quite different from gene replication.
Studies on information and cognitive sciences have provided accumulating evidence that such intelligent behavior often involves a stochastic data production process.
It has also been indicated that statistical parameters controlling the process are transmitted from individuals to individuals through statistical learning of the generated cultural products (e.g.\ \cite{Saffran1996,Fernandez2013,Lecun2015}).
This motivates us to study the signatures and consequences of statistical learning from the perspectives of cultural evolution and social learning \cite{Baxter2006,Griffiths2007,Reali2010,Perreault2012,Thompson2016,Chazelle2019}.

Signatures of the underlying dynamical process in complex biological and social systems are often presented as distribution laws that characterize the collective behavior of a system.
If found, these distribution laws can serve as nontrivial constraints in model building and facilitate the theoretical understanding of the process.
Examples of such distribution laws include Zipf’s law for the mobile dynamics of interacting individuals \cite{Marsili1998} and for chemical reaction dynamics in self-reproducing cells \cite{Furusawa2003}, and Taylor’s law for the random diffusion process of traffic networks \cite{Meloni2008}.

There are ongoing projects on the big-data analysis of cultural products \cite{Michel2011,Mauch2015,Elgammal2018}, and it has been found by analyzing classical music data that the frequencies of certain musical elements characterizing music styles exhibit approximate beta distributions \cite{Nakamura2019}.
The beta distribution is known in Bayesian statistics as an instance of conjugate distributions; a conjugate distribution is defined as a distribution that can parameterize both the prior and posterior distributions of some statistical model \cite{PRML}.
This observation suggests a connection between the distribution form of a cultural trait and the statistical learning process involved in the transmission of the trait.
However, a theoretical foundation for understanding this connection has not yet been established.

The purpose of this study is to quantitatively understand how the distribution of a cultural trait evolves through a transmission process involving statistical learning.
First, a number of statistical traits in music data are analyzed to extract empirical distribution laws.
We analyze various frequency statistics in music data created in different societies, which extends the observations in Ref.~\cite{Nakamura2019}.
In addition, global musical statistics such as the note density and scale of pitch intervals were analyzed.
These analyses suggest a general relationship between trait distribution and the underlying statistical model for cultural production.
Next, a model of cultural evolution incorporating statistical learning is studied and a relationship between the data production model and the equilibrium distribution of the corresponding trait is derived under simple and general assumptions regarding the transmission process.
The theoretical results explain the empirical distribution laws and extend them to a large class of cultural data production processes.
We also explore two applications of the present theory: (i) an efficient method for estimating trait distributions from small-size data and (ii) reproducing the evolution of some musical features in classical music data.

%%%%%%%%%%%%%%%%%%%%%%%%%%%%%%%%%%%%%%%%%%%%%%%%%%%%%%%
\section{Experimental observation}
\label{sec:ExperimentalObservation}
%%%%%%%%%%%%%%%%%%%%%%%%%%%%%%%%%%%%%%%%%%%%%%%%%%%%%%%

To extract empirical distribution laws in cultural data and, in particular, examine the ubiquity of the beta distribution law, we analyzed four datasets of music created in different societies and observed the distributions of frequencies of pitch and rhythm elements.
Musical pieces in these datasets are represented as symbolic musical scores in the musical instrument digital interface (MIDI) or MusicXML format, where the pitches of musical notes are represented as integers in units of semitones.
In a MusicXML file, one can also extract note durations (values) relative to a quarter note, as well as the barline information with which we can segment the note sequence into subsequences corresponding to measures.

The {\it classical} music dataset consists of pieces in various instrumentations \cite{Nakamura2019}, from which we can extract the pitch information.
All other datasets consist of melody data, from which we can extract information about pitches and note durations.
The {\it Wikifonia} dataset contains popular Western songs and jazz songs mainly composed in the early to mid-20th century \cite{Simonetta2018}.
The {\it J-pop} dataset contains popular songs created in Japan after 1950 \cite{JPopDataNobara}.
The {\it Irish} song dataset contains Irish folk songs collected from a public website~\cite{Sturm2016}.
All datasets contained $O(10^3$--$10^4)$ musical pieces created by various composers.
As cultural traits, we extracted the within-song frequencies of the intervals of consecutive pitches modulo 12 (pitch-class intervals; PCI) and those of the ratios of consecutive note durations (note-value ratios), and observed their distributions within each dataset.
See Appendix \ref{app:DataAnalysis} for details of the data and analysis method.

\begin{figure}
\centering
\includegraphics[width=0.95\columnwidth]{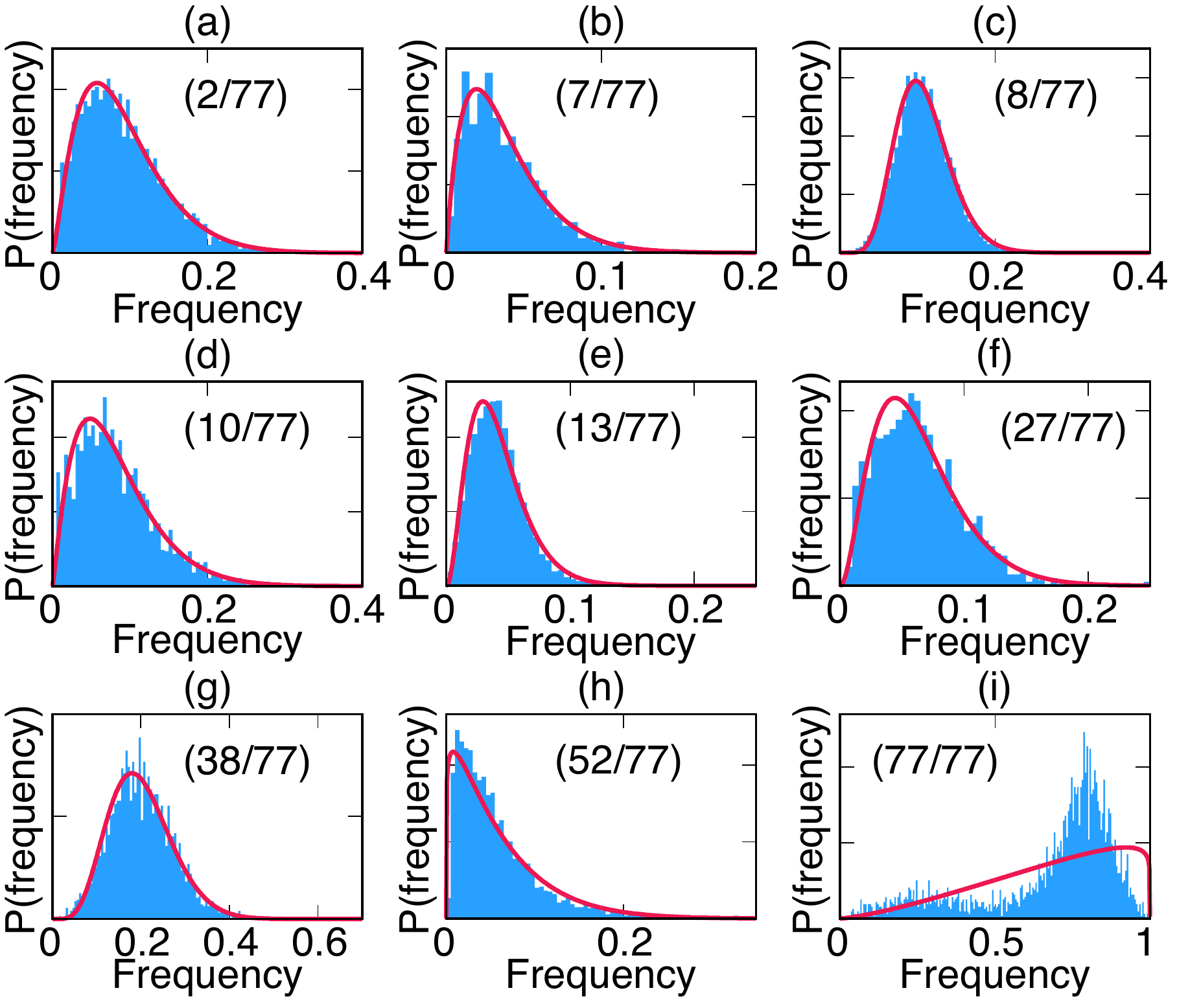}
\caption{Within-corpus distributions of within-song frequencies of pitch-class intervals (PCIs) and note-value ratios (NVRs) with fitted beta distributions. (a) Wikifonia PCI 9, (b) J-pop NVR 1:4, (c) Classical PCI 8, (d) Irish NVR 1:2, (e) Classical PCI 6, (f) J-pop PCI 5, (g) Irish PCI 10, (h) Wikifonia NVR 1:3, and (i) Irish NVR 1:1 (``PCI $i$'' represents PCIs of $i$ semitones). Numbers in panels indicate ranks of fitting quality (ordered in cumulative difference of data and fitted distributions).}
\label{fig:BetaDistributionData}
\vspace{-10pt}
\end{figure}

As shown in Fig.~\ref{fig:BetaDistributionData}, approximate beta distributions were observed for various musical elements across different datasets, which extensively generalize the previous observation of beta distributions in classical music data \cite{Nakamura2019}.
We note that the beta distributions have only two parameters; for example, when the parameters are adjusted to match the mean and the variance, the third moment or skewness can no longer be freely adjusted.
Two-thirds of the $77$ distributions analyzed had a fitting quality (measured by the cumulative difference of the data and fitted distributions) similar to or better than that of the example in Fig.~\ref{fig:BetaDistributionData}(h).
As exemplified in Fig.~\ref{fig:BetaDistributionData}(i), distributions that significantly deviated from the beta distribution were typically widespread and had multiple peaks corresponding to distinct music styles (e.g.\ duple rhythm vs dotted rhythm).

\begin{figure}
\centering
\includegraphics[width=0.65\columnwidth]{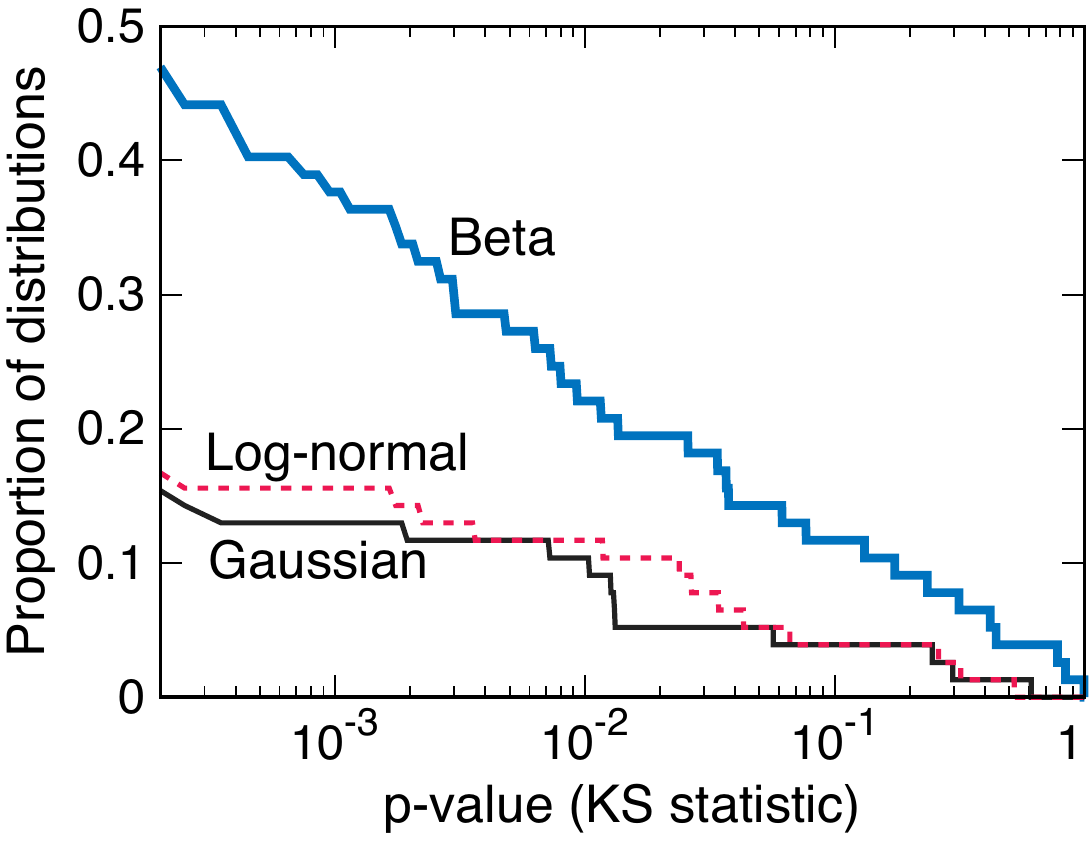}
\caption{Complementary cumulative distribution functions of p-values of KS statistics for fittings with beta, log-normal, and Gaussian distributions.}
\label{fig:KSPval}
\vspace{-10pt}
\end{figure}

To quantify the fitting quality more formally and compare it with alternative hypotheses, the distribution of the p-values of the Kolmogorov–Smirnov (KS) statistics is shown in Fig.~\ref{fig:KSPval}.
As alternative hypotheses, the log-normal and Gaussian distributions are compared in the figure; these distributions have two adjustable parameters, similar to the beta distribution.
The gamma distribution, which also has two parameters, is not compared here because it is a limiting distribution of the beta distribution for small values and fits the analyzed data equally well as the beta distribution in most cases.
Although the proportion of distributions that are statistically consistent with the beta distribution is not high ($14\%$ at the $95\%$ confidence level), the beta distribution has a considerably higher fitting quality than the other distributions.
Therefore, the data support that the beta distribution is the best approximation among the simple candidate distributions with two parameters.

\begin{figure}
\centering
\includegraphics[width=0.99\columnwidth]{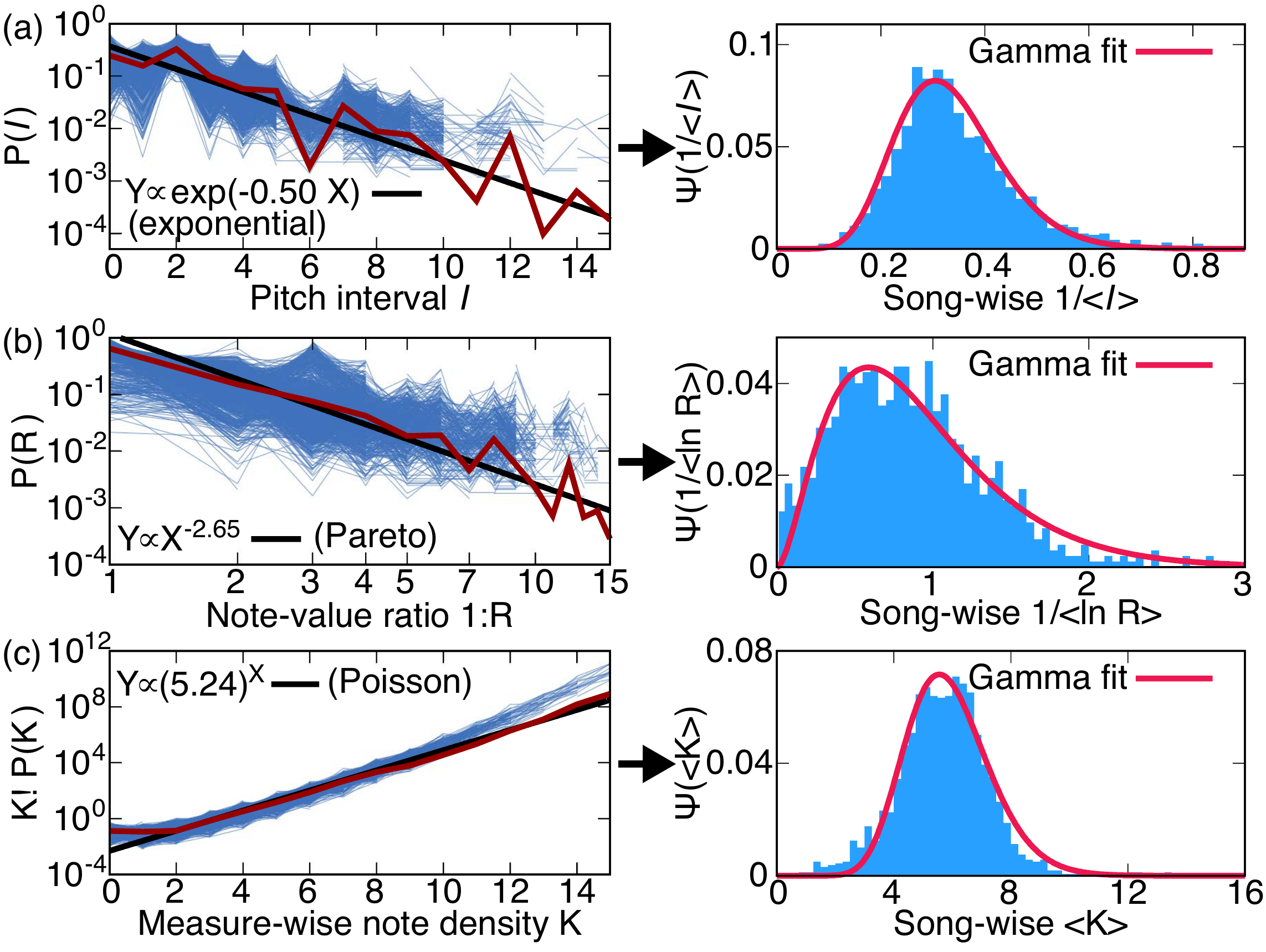}
\caption{Distributions of global musical statistics (J-pop data). (a) Left: Distributions $P$ of pitch intervals within songs (fine line segments), that within the dataset (bold line segments), and fitted exponential distribution (straight line). Right: Distribution $\Psi$ of song-wise statistics (reciprocals of means) of pitch intervals and fitted gamma distribution. (b) Similar results for note value ratios of the form $1$:$R$, where the global distribution is fitted by the Pareto distribution. (c) Similar results for measure-wise note densities, where the global distribution is fitted by the Poisson distribution.}
\label{fig:GlobalStatisticsJpop}
\end{figure}

We also analyzed song-level global statistics that obey certain statistical distributions using the same datasets.
Fig.~\ref{fig:GlobalStatisticsJpop} shows the results for the J-pop dataset.
First, we observe the distribution of pitch intervals $I$, which are defined as the absolute values of the differences between two consecutive pitches, within individual songs.
The overall contour of this distribution is approximated by an exponential distribution $P(I)\propto e^{-\lambda I}$ in the range $1\leq I\leq 12$, whereas the fine structure reflects the rarity of dissonant intervals (Fig.~\ref{fig:GlobalStatisticsJpop}(a)).
The rate parameter $\lambda$ of the exponential distribution can be estimated as $\lambda=1/\langle I\rangle$ for each song, where, in this section, $\langle\,\cdot\,\rangle$ denotes the average within a song.
This statistic represents the overall scale of the pitch intervals of the song.
The within-corpus distribution $\Psi$ of the rate parameters $\lambda$ of the individual songs approximately follows the gamma distribution.

Second, we observe the distribution of note value ratios $1$:$R$ of integer values $R$, where note value ratios are defined as the ratio of two consecutive note values, within individual songs.
The overall contour of this distribution can be approximated by a Pareto (power-law) distribution $P(R)\propto R^{-\alpha}$ in the range $1\leq R\leq10$.
The shape parameter $\alpha$ of the Pareto distribution can be estimated as $\alpha=1/\langle {\rm ln}\,R\rangle$ for each song.
This statistic represents the overall proportion of contrasted rhythms used in the song.
The within-corpus distribution $\Psi$ of the shape parameter $\alpha$ also roughly follows the gamma distribution (Fig.~\ref{fig:GlobalStatisticsJpop}(b)).

Last, we observe the distribution of the number of notes in a measure (measure-wise note density) $K$ within individual songs.
This distribution approximately obeys a Poisson distribution $P(K)\propto \rho^K/K!$ in the range $2\leq K\leq12$.
The rate parameter $\rho$ of the Poisson distribution can be estimated as $\rho=\langle K\rangle$ for each song.
This statistic represents the overall note density of the song.
The within-corpus distribution $\Psi$ of the rate parameters $\rho$ roughly follows the gamma distribution (Fig.~\ref{fig:GlobalStatisticsJpop}(c)).
The results for the other datasets are similar (see Appendix~\ref{app:DataAnalysis}).

An analysis using the KS statistics showed that the qualities of the gamma distribution fittings were lower than those of the beta distribution fittings for the frequency statistics.
Only one of the 10 fitted gamma distributions had a p-value greater than $5\%$.
Comparisons with the alternatives (log-normal and Gaussian distributions) indicated that the gamma distribution fits the data better than the others on average; it was the best-fitting distribution for 5 of the 10 samples.
Therefore, the global statistics data often suggest that the gamma distribution is the best approximation among simple choices, even though the fitting quality is not very high.

The observed gamma distributions are conjugate distributions for the exponential, Pareto, and Poisson distributions with respect to the analyzed parameters.
Together with the beta distribution law, the observations suggest a general conjugate relation between the within-product statistical distribution and the within-population distribution of the statistical cultural traits, which we call the {\it conjugate distribution law}.
The fact that this relation is observed in music data corresponding to various societies and time periods indicates an underlying general mechanism.

%%%%%%%%%%%%%%%%%%%%%%%%%%%%%%%%%%%%%%%%%%%%%%%%%%%%%%%
\section{Model}
\label{sec:Model}
%%%%%%%%%%%%%%%%%%%%%%%%%%%%%%%%%%%%%%%%%%%%%%%%%%%%%%%

%%%
\subsection{Formulation of dynamical system}
%%%

To explore a possible origin of the conjugate distributions, we consider an evolutionary model called the statistical learn-generate (SLG) system.
Suppose that at each generation $t$ there are $N$ individuals that generate cultural products $X^t_n$ ($n=1,\ldots,N$).
Each product (e.g.\ a musical piece) $X^t_n=(x^t_{n\ell})_{\ell=1}^L$ consists of $L$ samples (e.g.\ musical notes) $x^t_{n\ell}$ that are independently generated by a probability distribution $\phi(x;\theta^t_n)$.
This is called a data production model, where parameters $\theta^t_n$ are considered as cultural traits.
For simplicity, the set of all products created by an individual is treated as a single product of the same size $L$.
We assume that the distributions of all individuals are identical, but their parameter values can be different.
For example, if $\phi$ is a Bernoulli distribution, sample $x^t_{n\ell}$ takes $0$ or $1$, and $\theta$ is the probability of obtaining $1$.
Usually, $1$ represents the presence of a specific element we focus on (e.g.\ a PCI of 6 semitones), and $0$ represents the other cases.
When $\phi$ is Poisson distributed, sample $x^t_{n\ell}$ takes a nonnegative integer, and $\theta$ represents the rate.
Individual $n$ is often identified with its cultural trait $\theta^t_n$.

In general, we can incorporate a selection process for the individuals to produce cultural offspring in the next generation.
As the process and strength of selection of cultural products are still uncertain \cite{Salganik2006,Bentley2007}, we here consider a simple case without selection to focus on the effect of the transmission process.
In addition, each individual $\theta^{t+1}_n$ is assumed to have one dominant cultural parent (the primary parent) $\theta^{t}_n$.
For simplicity, each parent $\theta^t_n$ is assumed to have one cultural offspring $\theta^{t+1}_n$, which gives the same analytical result as the random selection case in the limit of an infinite population size.
It is likely in cultural transmission that products by other individuals (the secondary parents) in the parent generation will also be used for learning (oblique transmission \cite{CavalliSforzaFeldman}).
In this case, the set of products (training data) used for learning the cultural trait is composed of the primary parent's product and the products of the secondary parents.

\begin{figure}
\centering
\includegraphics[width=0.9\columnwidth]{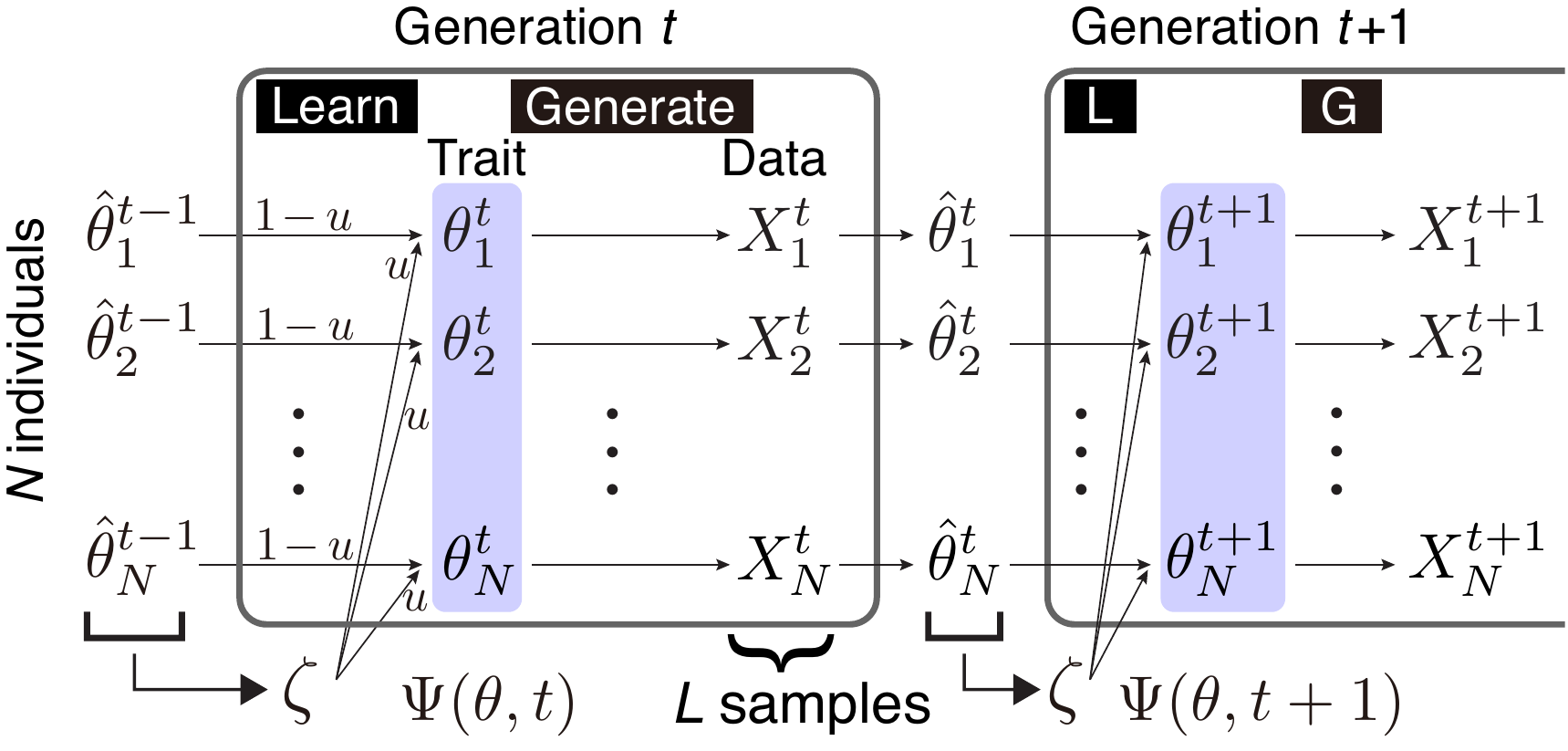}
\caption{Statistical learn-generate (SLG) system with oblique transmission.}
\label{fig:SLGSystemLinPres}
\vspace{-5pt}
\end{figure}

As will be validated later (below Eq.~(\ref{eq:ConjugateDistribution})), we assume that each parameter of $\theta^{t+1}_n$ is obtained by calculating the expectation value of some statistic extracted from the training data.
We denote the expectation value calculated from product $X^t_n$ as $\hat{\theta}^t_n$.
If the offspring use only their primary parent's product for learning (vertical transmission), then $\theta^{t+1}_n=\hat{\theta}^t_n$ holds.
When the products of secondary parents are also used, these products contribute to the calculation of the expectation value, and parameterization $\theta^{t+1}_n$ is given as a weighted sum of $\hat{\theta}^t_n$ and $\hat{\theta}^t_{n'}$ of the secondary parents $n'$.
If the secondary parents contribute equally, then the effect of oblique transmission can be represented as
\begin{equation}
\theta^{t+1}_n=(1-u)\hat{\theta}^t_n+u\zeta,\quad
\zeta=\frac{1}{N'}\sum_{n'}\hat{\theta}^t_{n'},
\label{eq:ObliqueTransmission}
\end{equation}
where the summation is taken over the secondary parents, $N'$ is their total number, and $u$ represents the strength of their influence (Fig.~\ref{fig:SLGSystemLinPres}).
When the secondary parents are randomly chosen and $N'$ is large, as assumed in the following analysis, we have $\zeta\simeq\bar{\theta}=(1/N)\sum_n\theta^t_n$ (population average), and $\zeta$ is effectively a constant.

When $\phi$ is Bernoulli distributed, the SLG system is equivalent to the (asexual) Wright--Fisher (WF) model used in population genetics \cite{Ewens}.
The neutrally selective WF model consists of a population of $N_g$ individuals, each having a gene with two alleles ($0$ and $1$).
In each generation, all individuals are replaced by new individuals, and the genes of the randomly chosen parents are inherited.
The gene frequency follows the same stochastic process as a vertically transmitted trait of the SLG system with $L=N_g$; thus, the SLG system is equivalent to a set of $N$ populations of the WF model.
Moreover, the effect of oblique transmission in Eq.~(\ref{eq:ObliqueTransmission}) with a constant $\zeta$ is equivalent to the linear pressure representing the effect of constant migration (from outside) or mutation \cite{Wright1931}.

%

%%%
\subsection{Equilibrium distribution}
%%%

We now consider the time evolution of the within-population distribution $\Psi(\theta,t)=P(\theta^t_n=\theta)$ of cultural traits in the large $N$ limit.
This problem is well understood for the WF model, and accordingly, for the case in which $\phi$ is Bernoulli distributed \cite{Crow1970}.
In the case of pure vertical transmission, the variable $\theta^t_n$ fluctuates statistically until it becomes $0$ or $1$, and the equilibrium distribution is given as $\Psi(\theta,\infty)=(1-\bar{\theta})\delta(\theta)+\bar{\theta}\,\delta(\theta-1)$.
In the presence of oblique transmission (linear pressure), a nontrivial equilibrium distribution was found by Wright \cite{Wright1931}.
To derive an analytical solution, we take the continuous time limit and a large $L$ limit.
Then, the dynamics can be described by the Fokker--Planck (FP) equation:
\begin{equation}
\frac{\partial\Psi(\theta,t)}{\partial t}=-\frac{\partial}{\partial\theta}\bigg\{M(\theta)\Psi(\theta,t)-\frac{1}{2}\frac{\partial[K(\theta)\Psi(\theta,t)]}{\partial\theta}\bigg\},
\label{eq:FP}
\end{equation}
where $M(\theta)=u(\zeta-\theta)$, $K(\theta)=\theta(1-\theta)/L$, and the terms in the curly brackets represent the probability current \cite{Crow1970}.
The zero-current equilibrium solution of the FP equation is given by the beta distribution ${\rm Beta}(\theta;2Lu\zeta,2Lu(1-\zeta))$ \cite{Wright1931}.
Therefore, the SLG system can explain the evolutionary origin of the beta distribution law in the presence of oblique transmission.

To extend the result for cultural traits following more general data production models, we consider exponential family (EF) distributions with one parameter $\theta$:
\begin{equation}
\phi(x;\theta)={\rm exp}\big[F(x)B(\theta)-A(\theta)+U(x)\big].
\label{eq:ExpFamily}
\end{equation}
A large class of probability distributions including Bernoulli, Poisson, Gaussian, and gamma distributions can be represented by different choices of $F(x)$ and $U(x)$.
Functions $B(\theta)$ and $A(\theta)$ specify how the distribution is parameterized.
We have $\langle F\rangle=A'(\theta)/B'(\theta)$ from $0=\partial_\theta\int dx\,\phi(x;\theta)$.
We can explicitly construct a conjugate distribution for the distribution in Eq.~(\ref{eq:ExpFamily}) as
\begin{equation}
\tilde{\phi}(\theta;\chi,\nu)={\rm exp}\big[\chi B(\theta)-\nu A(\theta)+C(\theta)+W(\chi,\nu)\big],
\label{eq:ConjugateDistribution}
\end{equation}
where $C(\theta)$ is a function, and $W(\chi,\nu)$ represents the normalization term.
The function $C$ does not change the conjugacy property and the minimal case with $C=0$ is usually used in Bayesian statistics.
To fix a parameterization scheme for $\theta$, we note that it is natural to use the expectation value $\langle F\rangle$ as a parameter.
It is known that the Cram\'{e}r--Rao bound can be attained only with this parameterization \cite{Fend1959}, meaning that it is the most efficient one for statistical learning.
Thus, we assume the relation $\langle F\rangle=A'(\theta)/B'(\theta)=\theta$.
The estimation variance of $F$ is given by $D(\theta)\equiv\langle(F-\theta)^2\rangle=1/B'(\theta)$, and distribution $\Psi(\theta,t)$ follows Eq.~(\ref{eq:FP}) with $K(\theta)=D(\theta)/L$.

With the oblique transmission effect $M(\theta)=u(\zeta-\theta)$, the zero-current equilibrium solution $\Psi_*(\theta)$ of Eq.~(\ref{eq:FP}) satisfies the following equation:
\begin{equation}
2Lu(\theta-\zeta)\Psi_*=-\frac{d}{d\theta}(D\Psi_*).
\label{eq:ZeroCurrentEquation}
\end{equation}
The solution of this equation can be written as
\begin{equation}
\Psi_*(\theta)\propto \frac{1}{D(\theta)}\,{\rm exp}\bigg[\int^\theta\frac{2Lu(\zeta-\eta)}{D(\eta)}d\eta\bigg].
\label{eq:ZeroCurrentSolutionIntegralForm}
\end{equation}
Using $D(\theta)^{-1}=B'(\theta)$ and $\theta=A'(\theta)/B'(\theta)$, we obtain
\begin{equation}
\Psi_*(\theta)\propto{\rm exp}\big[2Lu\zeta B(\theta)-2LuA(\theta)+{\rm ln}\,B'(\theta)\big].
\label{eq:ZeroCurrentSolution}
\end{equation}
Comparing with Eq.~(\ref{eq:ConjugateDistribution}), this is a conjugate distribution with $C={\rm ln}\,B'$, which shows the origin of the general conjugate distribution law in the presence of oblique transmission.
Eqs.~(\ref{eq:ZeroCurrentSolutionIntegralForm}) and (\ref{eq:ZeroCurrentSolution}) show how the equilibrium distribution form of cultural traits in the population depends on the statistical model for cultural production; $\Psi_*$ is essentially determined by the estimation variance $D$.

We can further say that the solution (\ref{eq:ZeroCurrentSolution}) is the minimal conjugate distribution if ${\rm ln}\,B'(\theta)$ is a linear combination of $B(\theta)$ and $A(\theta)$ up to a constant term:
${\rm ln}\,B'(\theta)=c_0+c_1B(\theta)+c_2A(\theta)$ for some constants $c_i$.
We obtain an equivalent condition by differentiating both sides as $B''(\theta)/B'(\theta)=(c_1+c_2\theta)B'(\theta)$, where we have used $A'(\theta)/B'(\theta)=\theta$.
Since $D(\theta)=1/B'(\theta)$, this is equivalent to the condition that $D'(\theta)$ is linear in $\theta$, or, $D(\theta)$ is quadratic in $\theta$.
Therefore, the equilibrium distribution is the minimal conjugate distribution if and only if the estimation variance $D(\theta)=1/B'(\theta)$ is quadratic:
\begin{equation}
D(\theta)=d_0+d_1\theta+d_2\theta^2,
\label{eq:QuadraticCondition}
\end{equation}
where $d_i$ are constants.
We can also confirm this result by substituting Eq.~(\ref{eq:ConjugateDistribution}) with $C=0$ into the zero-current equation, and obtain an explicit solution with
\begin{equation}
\chi=2Lu\zeta-d_1,\quad \nu=2Lu+2d_2.
\label{eq:ExplicitEquiSol}
\end{equation}

Table \ref{tab:Examples} lists examples of probability distributions that satisfy Eq.~(\ref{eq:QuadraticCondition}).
It can be shown that if $\phi(x;\theta)$ satisfies Eq.~(\ref{eq:QuadraticCondition}), so does the distribution for variable $y=f(x)$ obtained by a transformation function $f$ ($D$ remains unchanged).
For example, the Pareto distribution satisfies Eq.~(\ref{eq:QuadraticCondition}) with respect to the logarithmic mean statistic because the gamma distribution does with respect to the mean statistic.
EF distributions with quadratic estimation variances cover a wide range of well-known distributions \cite{Morris1982}, including those shown in Fig.~\ref{fig:GlobalStatisticsJpop} (see Appendix \ref{app:ExamplesOfQVDistributions}), indicating that the conjugate distributions found in cultural products are often minimal.
Not all EF distributions satisfy Eq.~(\ref{eq:QuadraticCondition}): a counterexample is the gamma distribution  with respect to the logarithmic mean statistic.

\begin{table}[t]
\centering
\tabcolsep = 5pt
\begin{tabular}{lll}
%\toprule
\bhline{1pt}
\begin{tabular}{@{\hspace{0em}}l}Individual's gener-\\ative model $\phi$\end{tabular} & \begin{tabular}{@{\hspace{0em}}l}Cultural\\trait $\theta$\end{tabular} & \begin{tabular}{@{\hspace{0em}}l}Trait distri-\\bution $\Psi$\end{tabular}
\\
\hline
%\midrule
Bernoulli & Mean & Beta
\\
Poisson & Mean & Gamma
\\
Gaussian & Mean & Gaussian
\\
Gaussian & Variance & Inverse gamma
\\
Gamma & Mean & Inverse gamma
\\
\bhline{1pt}
%\bottomrule
\end{tabular}
\caption{Examples of exponential family distributions satisfying Eq.~(\ref{eq:QuadraticCondition}) and their conjugate distributions.}
\label{tab:Examples}
\end{table}
%

%%%
\subsection{Properties of equilibrium distribution}
\label{sec:PropertiesOfEquilibriumDistribution}
%%%

Some properties of the equilibrium distribution can be derived from further analyses.
First, under a regular condition that is usually met, the relation $\langle\theta\rangle_*\equiv\int d\theta\,\theta\Psi_*(\theta)=\zeta$ holds (even when $\zeta$ is an external parameter).
This is consistent with the assumption of a constant population mean $\zeta=\bar{\theta}$.
Second, with the condition (\ref{eq:QuadraticCondition}), one can show that
\begin{equation}
V_*(\theta)=\int d\theta\,(\theta-\langle\theta\rangle_*)^2\Psi_*(\theta)=\frac{D(\zeta)}{2Lu-d_2},
\label{eq:VarianceAtEquilibrium}
\end{equation}
indicating that when individuals of the SLG system transmit different cultural traits $\theta$ simultaneously, the values $V(\theta)/D(\bar{\theta})$ at equilibrium  are independent of $\bar{\theta}$ and depend only on $Lu$ and $d_2$.
The derivation of these results is provided in Appendix \ref{app:AnalysisOfEquilibriumDistribution}.

\begin{figure}
\centering
\includegraphics[width=0.95\columnwidth]{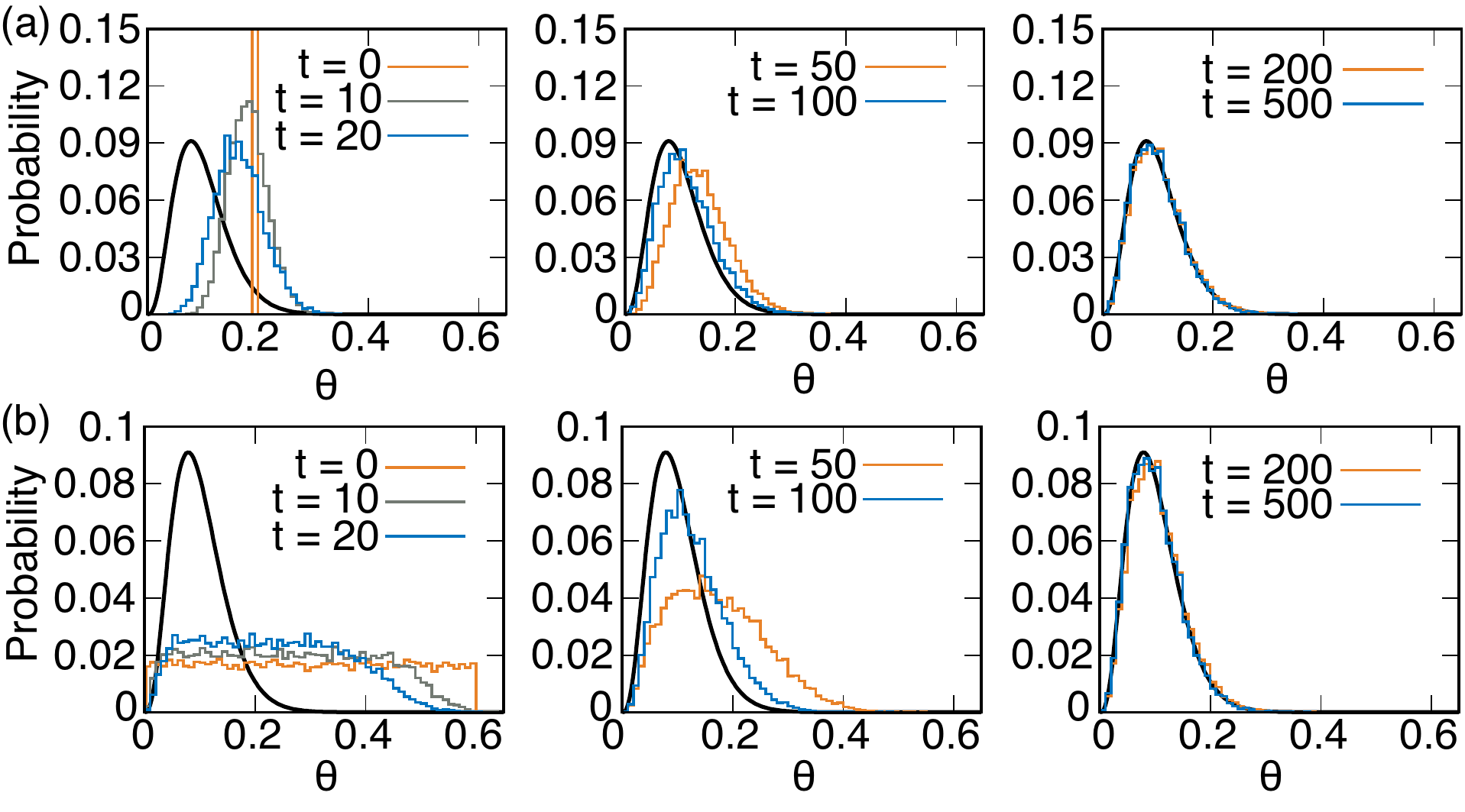}
\caption{Simulation results for SLG system with Bernoulli distributed production models. Parameters were set as $N=20000$, $L=1000$, $u=1/50=0.02$, and $\zeta=0.1$. Fitted distribution is ${\rm Beta}(\theta;4,36)$. (a) Initial distribution concentrated at $\theta=0.2$. (b) Initial distribution uniformly distributed in range $[0,0.6]$.}
\label{fig_Sim_BernoulliSLGS}
\end{figure}
Next, the relaxation time is of order $u^{-1}$, which is physically expected as the oblique transmission pushes a proportion $u$ of the trait toward $\zeta$ in unit time.
For the distributions listed in Table \ref{tab:Examples}, the relaxation time can be explicitly calculated by analyzing the time-dependent FP equation (\ref{eq:FP}) using the eigenfunction method (see Appendix \ref{app:AnalysisOfEquilibriumDistribution}).
This result is also confirmed by the numerical calculation in Fig.~\ref{fig_Sim_BernoulliSLGS}, whereby we conducted simulation experiments for the SLG systems with Bernoulli-distributed production models (with constant $\zeta$).
We see that the relaxation time is of the order $u^{-1}$ for the two extreme initial distributions.
The details of the simulation and results for other production models are given in Appendix \ref{app:NumericalAnalysisOfEquilibriumDistribution}.
This result indicates that the conjugate distribution can be maintained even under selective pressure as long as its effect is small on a time scale of order $u^{-1}$.

Without experimentally knowing the value of $u$, which is difficult at present, it is not possible to estimate the actual time necessary to reach equilibrium.
We can expect, however, that once the (quasi-)equilibrium distribution is realized in some culture in a society (e.g.\ Western classical music), a descendant culture in that society (e.g.\ Western popular music) is likely to reach equilibrium quickly, as far as many of the traditions are retained in the new culture.
For example, the existence of universal features in music worldwide indicates that many aspects of music are inherited over a long time \cite{Savage2015}.

In relating the theory with experiments, it is important to note that the creator's trait $\theta$ can only be observed through actual products.
When $\theta$ is estimated from a product with $L_p$ samples, a variance of $D(\theta)/L_p$ is expected, and the product-wise statistical parameters (as in Figs.~\ref{fig:BetaDistributionData} and \ref{fig:GlobalStatisticsJpop}) will have a distribution $\Psi_p$ distorted from that of $\theta$.
Nonetheless, since this distortion process is same as the transmission process, $\Psi_p$ approximately follows the same conjugate distribution as $\Psi_*$ with a variance increased from Eq.~(\ref{eq:VarianceAtEquilibrium}) by $D(\zeta)/L_p$.
This explains the empirical conjugate distribution laws observed in Figs.~\ref{fig:BetaDistributionData} and \ref{fig:GlobalStatisticsJpop}.
This argument also indicates the difficulty of applying Eq.~(\ref{eq:VarianceAtEquilibrium}) for estimating the value of $Lu$ directly from product data if $Lu\gg L_p$.
The values of $D(\bar{\theta})/V(\theta)$ obtained from the distributions in Fig.~\ref{fig:BetaDistributionData} were $29.5({\rm mean})\pm9.11({\rm s.d.})$ for the Wikifonia data (excluding distributions poorly fitted by beta distributions), which are relatively consistent among different musical elements and significantly smaller than the mean $\overline{L}_p=107$.
Since an extremely small $u$ is unlikely, these small values probably reflect statistical dependence of the samples due to frequent repetitions in music.
Similar results were obtained for the other data (see Appendix \ref{app:DataAnalysis}).

%%%
\subsection{Generalization for multi-parameter case}
\label{sec:FPEqMultiParameters}
%%%

We can extend the SLG system for EF distributions with more than one parameter and show that a conjugate distribution emerges at equilibrium in the presence of oblique transmission.
An exponential family distribution with multiple parameters $\theta=(\theta_i)_{i=1}^K$ is given as
\begin{equation}
\phi(x;\theta)={\rm exp}\bigg[\sum^K_{k=1}F_k(x)B_k(\theta)-A(\theta)+U(x)\bigg],
\end{equation}
where $F_k$ are independent statistics and $x=(x_k)_{k=1}^{K'}$ denotes a set of stochastic variables.
In the following, we write $\partial_i=(\partial/\partial \theta_i)$ and omit the summation symbol for paired indices.
From $\langle\partial_i\phi\rangle=0$, we obtain $\partial_iA=\langle F_k\rangle\partial_iB_k=G_{ik}\langle F_k\rangle$, where $G_{ik}\equiv\partial_iB_k$.
We assume that $G_{ik}$ is invertible: $G^{-1}_{\ell i}G_{ik}=\delta_{\ell k}$ and $G_{ik}G^{-1}_{kj}=\delta_{ij}$.
In the expectation value parameterization, the following equations hold:
\begin{equation}
\langle F_k\rangle=G^{-1}_{ki}\partial_iA=\theta_k,\quad \partial_iA=G_{ik}\theta_k.
\label{eq:MultiParamEVParameterization}
\end{equation}
From $\langle\partial_i\partial_j\phi\rangle=0$, after some calculation, we obtain the estimation covariance matrix as
\begin{equation}
D_{k\ell}\equiv\langle (F_k-\theta_k)(F_\ell-\theta_\ell)\rangle=G^{-1}_{k\ell}.
\label{eq:EstimationCovariance}
\end{equation}
Since the estimation covariance matrix $D_{k\ell}$ is symmetric, so is $G_{k\ell}$ in the expectation value parameterization.

The FP equation for the case with multiple parameters is given as
\begin{equation}
\frac{\partial\Psi(\theta,t)}{\partial t}=-\frac{\partial}{\partial\theta_i}\bigg[M_i(\theta)\Psi(\theta,t)-\frac{1}{2}\frac{\partial\{K_{ij}(\theta)\Psi(\theta,t)\}}{\partial\theta_j}\bigg].
\end{equation}
By substituting $M_i(\theta)=u(\zeta_i-\theta_i)$ (the oblique transmission effect) and $K_{ij}(\theta)=D_{ij}/L$, we obtain the zero-current equation as
\begin{equation}
2Lu(\zeta_i-\theta_i)\Psi-(\partial_jD_{ij})\Psi-D_{ij}\partial_j\Psi=0.
\label{eq:ZeroCurrentEqMultiParam}
\end{equation}
We assume the following conjugate distribution: $\Psi\propto{\rm exp}[\chi_kB_k(\theta)-\nu A(\theta)+C(\theta)]$.
This is a solution of Eq.~(\ref{eq:ZeroCurrentEqMultiParam}) if the following equations are satisfied:
\begin{equation}
\nu=2Lu,\quad \chi_i=2Lu\zeta_i,\quad \partial_jD_{ij}+D_{ij}\partial_jC=0.
\end{equation}
After some calculation, we find that the last equation is equivalent to $\partial_kC=G^{-1}_{ij}\partial_kG_{ji}$, which can be solved as $C={\rm ln}\,{\rm det}\,G$ using Jacobi's formula.
Therefore, the equilibrium solution is given as
\begin{equation}
\Psi_*\propto{\rm exp}[2Lu\zeta_kB_k(\theta)-2Lu A(\theta)+{\rm ln}\,{\rm det}\,G(\theta)].
\label{eq:EquivSolutionMultiParam}
\end{equation}
Details of the derivation are given in Appendix \ref{app:FPEqMultiParameters}.

As an example, we consider a discrete (categorical) distribution.
A $(K+1)$-dimensional discrete distribution is described by a variable $x=(x_i)_{i=1}^{K+1}$, where $x_i$ is either $0$ or $1$ and $\sum_{i=1}^{K+1}x_i=1$ (one-hot vector).
The parameters form a probability vector $\theta=(\theta_i)_{i=1}^{K+1}$ $(\sum_{i=1}^{K+1}\theta_i=1$).
Because of the normalization conditions, the last element can be represented by the other elements as $x_{K+1}=1-\sum_{i=1}^{K}x_i$ and $\theta_{K+1}=1-\sum_{i=1}^K\theta_i$, and the parameter space is in fact $K$-dimensional.
The data production model is an exponential family distribution:
\begin{equation*}
\phi(x;\theta)=\prod_{i=1}^{K+1}\theta_i^{x_i}={\rm exp}\bigg[\sum_{i=1}^Kx_i\{{\rm ln}\,\theta_i-{\rm ln}\,\theta_{K+1}\}+{\rm ln}\,\theta_{K+1}\bigg].
\end{equation*}
We have $F_i=x_i$, $B_i={\rm ln}\,\theta_i-{\rm ln}\,\theta_{K+1}$, and $A=-{\rm ln}\,\theta_{K+1}$, and thus $G_{ij}=\delta_{ij}/\theta_i+1/\theta_{K+1}$ and $\partial_iA=1/\theta_{K+1}$.
The estimation covariance matrix is $D_{ij}=G^{-1}_{ij}=\theta_i(\delta_{ij}-\theta_j)$.
We can show ${\rm det}\,G=({\rm det}\,D)^{-1}=(\theta_1\cdots\theta_{K+1})^{-1}$ by mathematical induction on $K$.
Substituting this into Eq.~(\ref{eq:EquivSolutionMultiParam}), we obtain
\begin{align*}
\Psi_*(\theta)\propto{\rm exp}\bigg[\sum_{i=1}^{K+1}(2Lu\zeta_i-1){\rm ln}\,\theta_i\bigg],
\end{align*}
where $\zeta_{K+1}\equiv 1-\zeta_1-\cdots-\zeta_K$.
Therefore, the equilibrium distribution is a Dirichlet distribution ${\rm Dir}(\theta;\bm\alpha)$ with Dirichlet parameters $\alpha_i=2Lu\zeta_i$ ($i=1,\ldots,K+1$).
It is not easy to find other examples where the minimal conjugate distribution is the equilibrium solution due to the non-diagonal elements of the estimation covariance matrix.

%%%%%%%%%%%%%%%%%%%%%%%%%%%%%%%%%%%%%%%%%%%%%%%%%%%%%%%
\section{Applications}
\label{sec:Application}
%%%%%%%%%%%%%%%%%%%%%%%%%%%%%%%%%%%%%%%%%%%%%%%%%%%%%%%

Here, we explore two applications of the theory developed in Sec.~\ref{sec:Model}.
In the first application, we examine how the conjugate distribution law can be used to efficiently estimate trait distributions from small-sized data.
The details of trait distributions are important inputs for building and testing models of evolution, and it becomes a challenging task when there are only a limited number of cultural products available for analysis.
In such a case, prior knowledge about the approximate form of trait distributions can be a useful guide as it can effectively reduce the number of parameters that need to be inferred from the data.
\begin{figure}
\centering
\includegraphics[width=0.99\columnwidth]{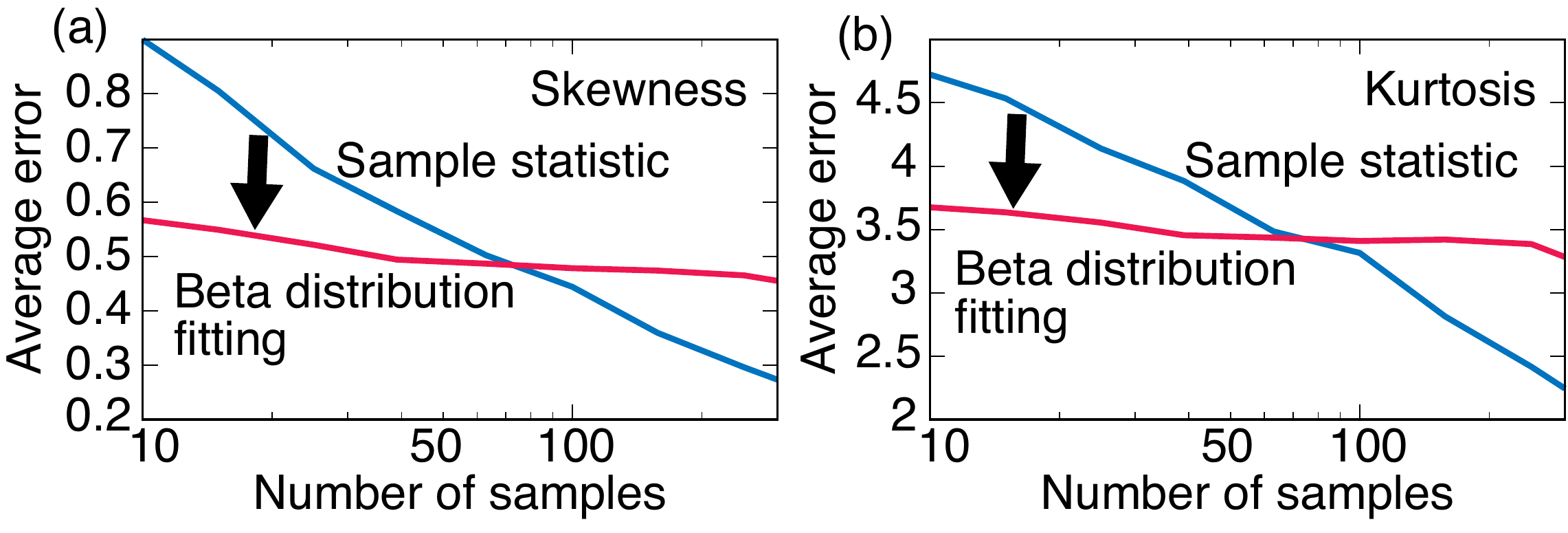}
\caption{Estimation errors of skewness and kurtosis.}
\label{fig:SummaryStatisticsErrorsForSmallData}
%\vspace{-10pt}
\end{figure}

As a specific problem, we used the data from the 77 frequency statistics analyzed in Fig.~\ref{fig:BetaDistributionData} and considered the estimation of skewness and kurtosis for each trait distribution.
These are the most important characteristics of a probability distribution next to the mean and variance.
If there is no prior knowledge about the trait distribution, then these four quantities (mean, variance, skewness, and kurtosis) should be estimated independently because we cannot assume any relationship between any two of them.
A standard method for estimating the skewness and kurtosis is to use the sample statistics for the third and fourth moments.
If the traits follow the conjugate distribution (the beta distribution in this case), however, then the trait distribution has only two independent parameters and after fitting these parameters, the skewness and kurtosis (and in fact all other moments) can be analytically calculated.
Specifically, we use the moment matching method, where the beta distribution parameters are determined by the sample mean and variance.

To examine the efficiency of the method using beta distribution fitting for small-size data, we simulated the reduction of data size by randomly constructing a subset of samples and compared the estimated values of the skewness and kurtosis using the reduced data and those using the original data.
To obtain the estimation errors, the following procedure was applied for each dataset and for each particular reduced number of samples: iterate 10 times the random data reduction and calculation of the absolute deviations of the estimated statistics from the sample statistics computed by using all of the data, and average the deviations.
In Fig.~\ref{fig:SummaryStatisticsErrorsForSmallData}, the average estimation errors for the standard method and the method using beta distribution fitting are plotted for varying numbers of samples.
As expected, the latter method becomes more efficient as the number of samples decreases, and for the music data analyzed, the method was more efficient when the number of samples was less than 70, for both the estimations of the skewness and kurtosis.
The same method can be applied when the trait distribution approximately follows the conjugate distribution.
\begin{figure}
\centering
\includegraphics[width=0.6\columnwidth]{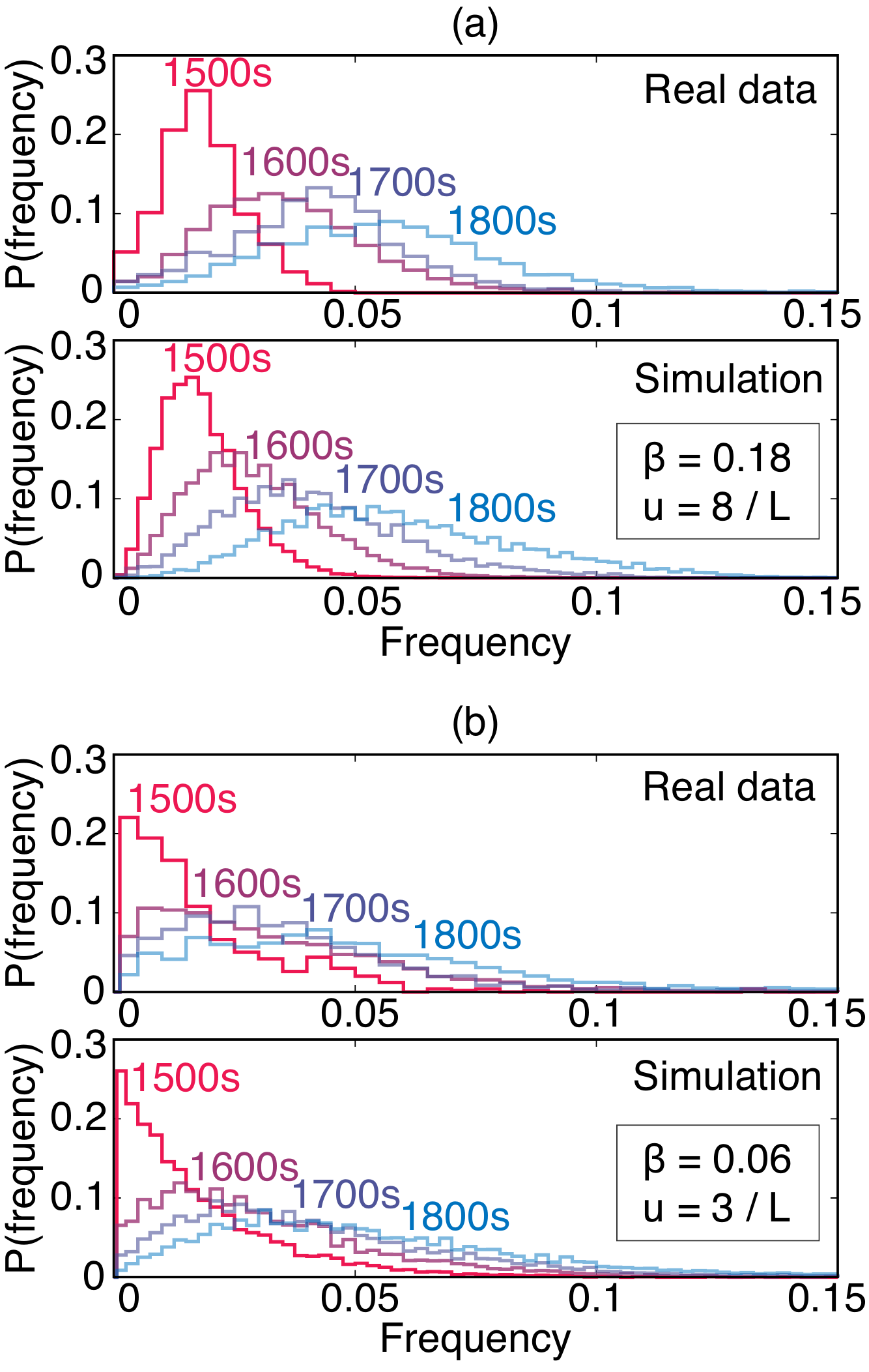}
\caption{Evolution of frequency distribution of (a) tritones and (b) nondiatonic motions in classical music dataset. Real data and simulation results.}
\label{fig:SLGSForClassicalMusicData}
%\vspace{-10pt}
\end{figure}

In the second application, we used the SLG system to reproduce the evolution of musical features in classical music data.
Ref.~\cite{Nakamura2019} analyzed the evolution of the distributions of the frequencies of two musical elements that represent two major aspects of tonal music, tritones and nondiatonic motions.
As reproduced in Fig.~\ref{fig:SLGSForClassicalMusicData}, the dynamics of these features have common patterns: the features are approximately beta distributed in each time period, and the mean and variance evolve significantly during the time period analyzed.
A model was developed in that study to account for the observed data, assuming that the traits follow the beta distribution at every time during the evolution and that selection pressures depending on the trait distribution drive the evolutionary process \cite{Nakamura2019}.

We use the SLG system to reproduce dynamic patterns, without manually imposing beta distributions in the dynamics.
We assume a simple (Wrightian) fitness with a log potential: $w(\theta)\propto e^{\beta\,{\rm ln}\,\theta}$.
A Monte Carlo simulation was conducted to compute the predictions of the model, and the parameters $\beta$ and $u$ were optimized to best fit the data.

The simulation results are shown in Fig.~\ref{fig:SLGSForClassicalMusicData}, where we set $N=10,000$ and $L=3,000$, and the initial populations were sampled from the beta distributions fitted to the initial distributions (in the 1500s).
The time unit was taken to be 10 years.
The optimal values of $\beta$ and $u$ found by a grid search in the ranges $\beta\in[0,1]$ and $u\in[0,50/L]$ are shown in the figures.
For both data, the optimal value of $u$ was positive, and the SLG system reproduced the overall pattern of evolution.
Therefore, the SLG system serves as a candidate model for explaining the evolution of classical music, an alternative to the model in Ref.~\cite{Nakamura2019}.
Although it is difficult to determine which is more appropriate solely from these data, the present model has the advantage that it does not constrain the distribution form for the traits by hand.

%%%%%%%%%%%%%%%%%%%%%%%%%%%%%%%%%%%%%%%%%%%%%%%%%%%%%%%
\section{Discussion}
%%%%%%%%%%%%%%%%%%%%%%%%%%%%%%%%%%%%%%%%%%%%%%%%%%%%%%%

We have shown that the distribution of cultural traits transmitted through statistical learning converges to an equilibrium distribution determined by the estimation variance of the trait, when oblique transmission is present and selective pressure is absent.
The equilibrium distribution becomes the conjugate distribution of the individuals' data production model when it is an exponential family (EF) distribution, and quantitatively depends on the product $Lu$ of the sample size $L$ of cultural products used for learning and the strength $u$ of oblique transmission.
The derived distribution forms were supported by empirical distributions extracted from music data.

In the present model, the existence of one primary cultural parent and a number of secondary parents has been assumed for simplicity, where the influence of the secondary parents was treated as a ``mean field.''
We can extend Eq.~(\ref{eq:ObliqueTransmission}) to a case with multiple primary parents by replacing the term $\hat{\theta}^t_n$ with the arithmetic mean of the expectation values of their products.
If the primary parents are randomly chosen, then the model is similar to the blending model of a continuous trait \cite{CavalliSforzaFeldman}, and we can roughly estimate the variance of the trait distribution at equilibrium as $O(D(\bar{\theta})/L)$.
In such a case, oblique transmission is expected to have a small effect when $u$ is small.
By contrast, when there is a tendency to choose individuals with similar traits as primary parents, the multi-parent model becomes similar to the one-parent model, and the effect of oblique transmission becomes more relevant.
In general, we can consider a network of influences, similar to Ref.~\cite{Baxter2006}, and extend the weights in the sum of Eq.~(\ref{eq:ObliqueTransmission}) to the distribution over the network.
It would be interesting to systematically study how the impact of oblique transmission changes according to the network structure using such models.

When selectively neutral, the asymptotic behavior of the SLG system is similar to that of iterated learning models with Bayesian agents \cite{Griffiths2007,Thompson2016}, in which the prior distribution is maintained at equilibrium.
This suggests the necessity of studying the dynamics under selection to discriminate between the different transmission mechanisms.
From the model-building perspective, the SLG system removes reliance on prior distributions and allows for flexible extensions with selective pressures.

Our findings indicate a significant link between statistical learning and cultural evolution, and connect them with statistical theory in an intriguing way.
Eqs.~(\ref{eq:ZeroCurrentSolution}) and (\ref{eq:QuadraticCondition}) indicate a nontrivial relationship between oblique transmission and Bayesian learning that reflects the geometric structure of the data production models \cite{Amari2016}.
Many statistical theories, including Bayesian estimation theory and limit theorems, can be developed for EF distributions that satisfy Eq.~(\ref{eq:QuadraticCondition}) \cite{Morris1982,Morris1983} and can now be applied to analyze cultural traits of a statistical nature.
Importantly, our results provide a theoretical background for choosing suitable parametric models for cultural traits.
For example, the widely applied Gaussian assumption \cite{CavalliSforzaFeldman,BoydRicherson} can be an inappropriate approximation.

As demonstrated in Sec.~\ref{sec:Application}, the SLG system opens new possibilities and questions for studying the dynamic aspects of culture and intelligence.
For example, the dynamic relation $\mu_t\propto\sigma_t$ between the mean $\mu_t$ and standard deviation $\sigma_t$ observed in classical music data \cite{Nakamura2019}, as opposed to $\sigma_t\propto\sqrt{\mu_t}$ expected from Eq.~(\ref{eq:VarianceAtEquilibrium}), implies that if the process is in a quasi-equilibrium state, then the factor $Lu$ is time-varying, similar to an evolving mutation rate \cite{Kaneko1992,Earl2004}.
Finally, it will be interesting to apply the presented analysis method to other cultural domains such as language \cite{Griffiths2004,Michel2011}, fine art \cite{Elgammal2018}, and cooking \cite{Caraher1999}, and to investigate the universal properties of cultural evolution.

%%%%%%%%%%%%%%%%%%%%%%%%%%%%%%%%%%%%%%%%%%%%%%
\section*{Acknowledgments}
%%%%%%%%%%%%%%%%%%%%%%%%%%%%%%%%%%%%%%%%%%%%%%
The author would like to thank Kunihiko Kaneko and Nobuto Takeuchi for useful discussions and Madison S.~Krieger for helpful comments on the manuscript.
This work was in part supported by JSPS KAKENHI No.\ 19K20340 and ISHIZUE 2021 of Kyoto University Research Development Program.

\appendix

%%%%%%%%%%%%%%%%%%%%%%%%%%%%%%%%%%%%%%%%%%%%%%
\section{Data analysis: Method and supplemental results}
\label{app:DataAnalysis}
%%%%%%%%%%%%%%%%%%%%%%%%%%%%%%%%%%%%%%%%%%%%%%

%%%%%%%%%%%%%%%%%%%%%%%%%%%%%%%%%%%%%%%%%%%%%%
\subsection{Datasets}
%%%%%%%%%%%%%%%%%%%%%%%%%%%%%%%%%%%%%%%%%%%%%%

For the analysis in Sec.~\ref{sec:ExperimentalObservation} we used four music datasets.
The {\it classical} music dataset consists of 9,727 musical pieces in various instrumentations written by various composers \cite{Nakamura2019}.
The data contents are represented in the MIDI format, and we extracted integer pitches in units of semitones for the analysis.
We did not use the rhythm information since it is unreliable due to the nature of the MIDI format.
The musical notes are ordered according to the onset time and the musical instruments in each file and we obtained the sequence of pitches in this order.
The {\it Wikifonia} dataset contains Western popular songs and jazz songs mainly composed in the early to mid 20th century and was compiled in Ref.~\cite{Simonetta2018}.
The dataset contains 6,388 songs (mostly vocal melodies).
The {\it J-pop} dataset contains 1,596 popular songs in Japan composed between 1946 and 2010 whose transcriptions are published in Ref.~\cite{JPopDataNobara}.
The {\it Irish} song dataset contains Irish folk songs collected from a public website and was compiled in Ref.~\cite{Sturm2016}.
6,345 songs in G major key and in 4/4 time were used for the analysis.
The contents of the Wikifonia, J-pop, and Irish song datasets are represented in the MusicXML format and we extracted the pitch and rhythm information.
The pitch information was extracted as a sequence of integer pitches.
The rhythm information was extracted as sequences of onset and offset times, both represented in beat units.
The score-notated duration (note value) of a musical note was defined as the difference between its onset and offset times.

%%%%%%%%%%%%%%%%%%%%%%%%%%%%%%%%%%%%%%%%%%%%%%
\subsection{Analysis of frequency statistics}
%%%%%%%%%%%%%%%%%%%%%%%%%%%%%%%%%%%%%%%%%%%%%%

For the analysis in Fig.~\ref{fig:BetaDistributionData}, we extracted the within-song frequencies of the intervals of consecutive pitches modulo 12 (pitch-class intervals; PCI) and those of the ratios of consecutive note durations (note-value ratios; NVR).
Given a sequence of pitches $(p_n)_{n=1}^N$, the sequence of PCI $(q_n)_{n=1}^{N-1}$ is obtained by $q_n\equiv p_{n+1}-p_n~({\rm mod}\,12)$, where $0\leq q_n\leq 11$.
For each PCI $1\leq q\leq 11$, the within-song PCI frequency was calculated by dividing the number of appearance of $q$ by the total number of musical notes in a song.
Following Ref.~\cite{Nakamura2019}, we did not use the zero PCI for the analysis since it describes a continuation of the same tone (up to octave equivalence) and has little information.
The PCI probabilities are independent of musical key and commonly used for analyzing music styles.
The within-song frequencies of NVRs are similarly defined by counting the number of times the ratio $r_n$:$r_{n+1}$ of consecutive note values is a specific ratio.
We calculated the frequencies of the most common ratios: 1:1, 1:2, 2:1, 1:3, 3:1, 2:3, 3:2, 1:4, 4:1, 1:6, and 6:1 (11 statistics in total).

The within-corpus distribution of the within-song frequencies was obtained for each of the four datasets.
The zero frequency samples were not used in the analysis of these distribution, as in Ref.~\cite{Nakamura2019}, since those samples can be contaminated with different styles/genres of music.
Since the 11 PCI statistics were calculated for all the datasets and the 11 NVR statistics were calculated for the datasets other than the classical music dataset, we analyzed 77 distributions in total.
\begin{figure}
\centering
\includegraphics[width=0.9\columnwidth]{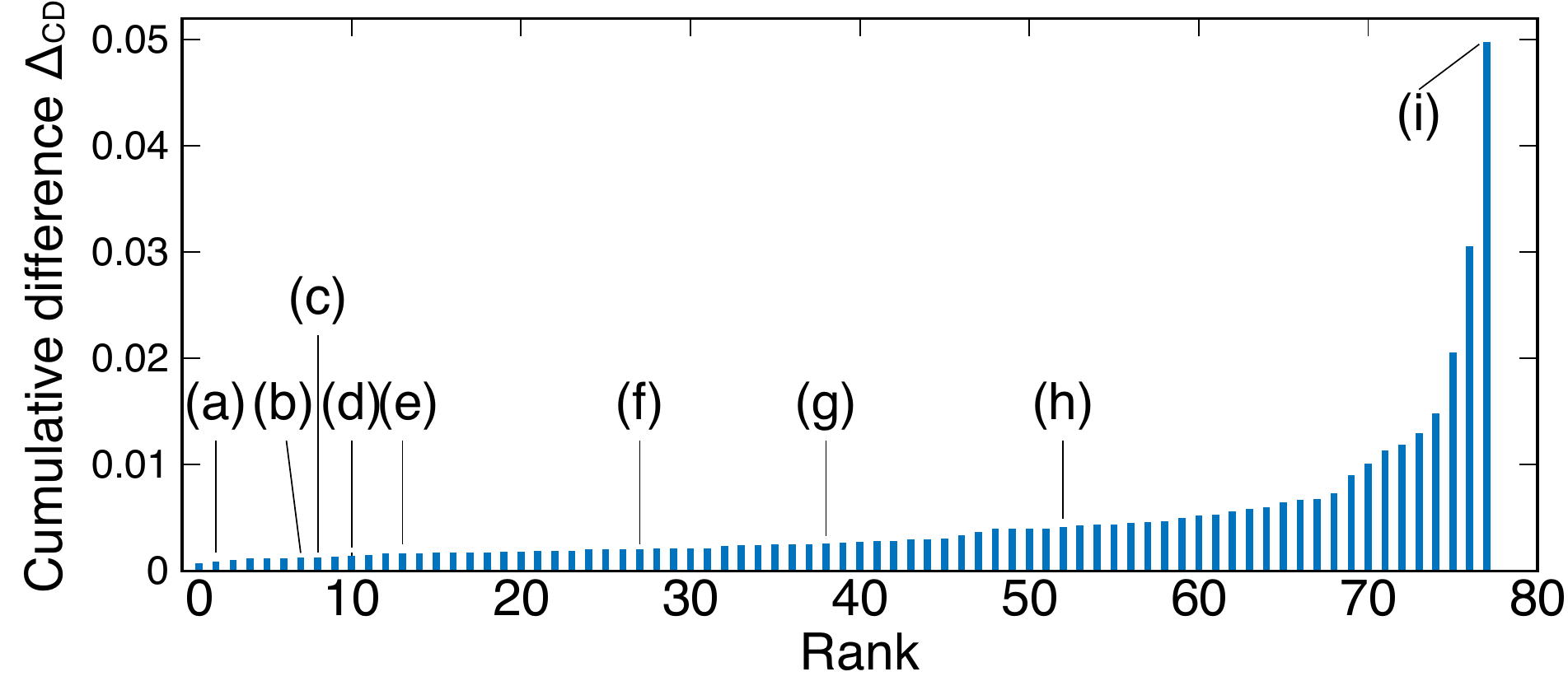}
\caption{Cumulative differences of the data and fitted beta distributions (representing the ``goodness of fit'') sorted by rank. The alphabetical labels represent the corresponding panels in Fig.~\ref{fig:BetaDistributionData}.}
\label{fig:CumDiff}
\end{figure}

For each distribution, the beta distribution was fitted by the moment-matching method using the mean and variance.
To quantitatively measure the ``goodness of fit'', we computed the cumulative difference $\Delta_{\rm CD}$ of the data and fitted beta distributions defined as
\begin{equation}
\Delta_{\rm CD}=\int d\theta\,|C_{\rm data}(\theta)-C_{\rm fit}(\theta)|,
\label{eq:CumDiff}
\end{equation}
where $C_{\rm data}(\theta)$ denotes the data cumulative distribution function (CDF) and $C_{\rm fit}(\theta)$ denotes the CDF of the fitted beta distribution.
In numerical analysis, the data distribution was represented by a histogram and the integral in Eq.~(\ref{eq:CumDiff}) was approximated by a summation.
The CDF was used here instead of the probability distribution function to reduce the dependence on the histogram bin width ($0.005$ in our analysis).
The distribution of the values of $\Delta_{\rm CD}$ is shown in Fig.~\ref{fig:CumDiff}.

For the estimation of $D(\bar{\theta})/V(\theta)$ discussed in Sec.~\ref{sec:PropertiesOfEquilibriumDistribution}, we used the distributions with $\Delta_{\rm CD}<0.004$ that included 51 (about two thirds) of the 77 distributions.
The (min, mean, max, s.d.) were ($40.5,58.3,76.1,13.1$) (classical), ($17.1,29.5,46.6,9.11$) (Wikifonia), ($39.3,47.5,57.4,5.0$) (J-pop), and ($24.7,39.0,61.6,11.0$) (Irish).
The average numbers of notes $\overline{L}_p$ within each corpus were $2300$ (classical), $107$ (Wikifonia), $148$ (J-pop), and $115$ (Irish).
For all datasets, the values of $D(\bar{\theta})/V(\theta)$ were similar among different statistics and significantly smaller than $\overline{L}_p$.

%%%%%%%%%%%%%%%%%%%%%%%%%%%%%%%%%%%%%%%%%%%%%%
\subsection{Analysis of global statistics}
%%%%%%%%%%%%%%%%%%%%%%%%%%%%%%%%%%%%%%%%%%%%%%

%
\begin{figure}
\centering
\includegraphics[width=0.95\columnwidth]{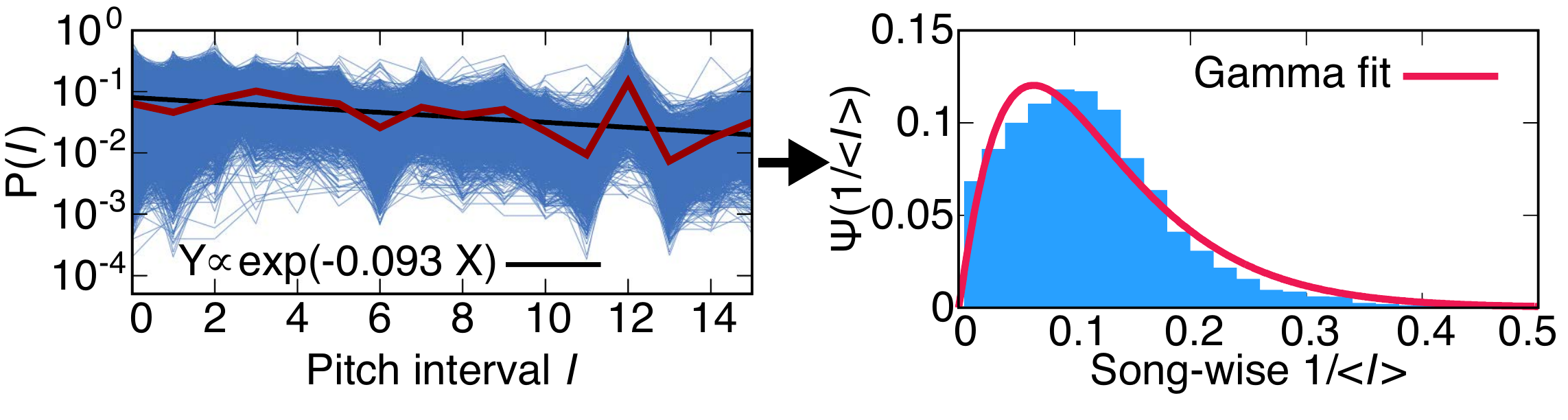}
\caption{Distributions of pitch-interval statistics in the classical music data. The left panel shows the distributions of pitch intervals within songs (fine line segments), that within the whole dataset (bold line segments), and a fitted exponential distribution (straight line); the right panel shows the distribution of song-wise statistics (reciprocals of the means) of pitch intervals and a fitted gamma distribution.}
\label{fig:GlobalStatisticsClassical}
\end{figure}
For the analysis in Fig.~\ref{fig:GlobalStatisticsJpop}, we calculated the song-wise global statistics as follows.
The probabilities of pitch intervals $I$ within songs were obtained by calculating the relative frequencies of pitch intervals $|p_{n+1}-p_n|$.
The rate parameter of the exponential distributions used to fit the contours of these distributions were calculated by the probability values in the range $1\leq I\leq12$.
To deal with missing samples and deviations due to the rareness of dissonant intervals, we used the method of least squares (the linear regression was applied in the logarithmic domain of $I$).
Similarly, the contour of the distribution of note-value ratios $1$:$R$ was fitted by a Pareto distribution in the range $1\leq R\leq 10$ by applying the linear regression in the log-log domain.
The probability distribution $P(K)$ of the measure-wise note density $K$ was fitted by a Poisson distribution by applying the linear regression for ${\rm ln}[K!P(K)]$ in the range $2\leq K\leq12$.
\begin{figure}
\centering
\includegraphics[width=0.95\columnwidth]{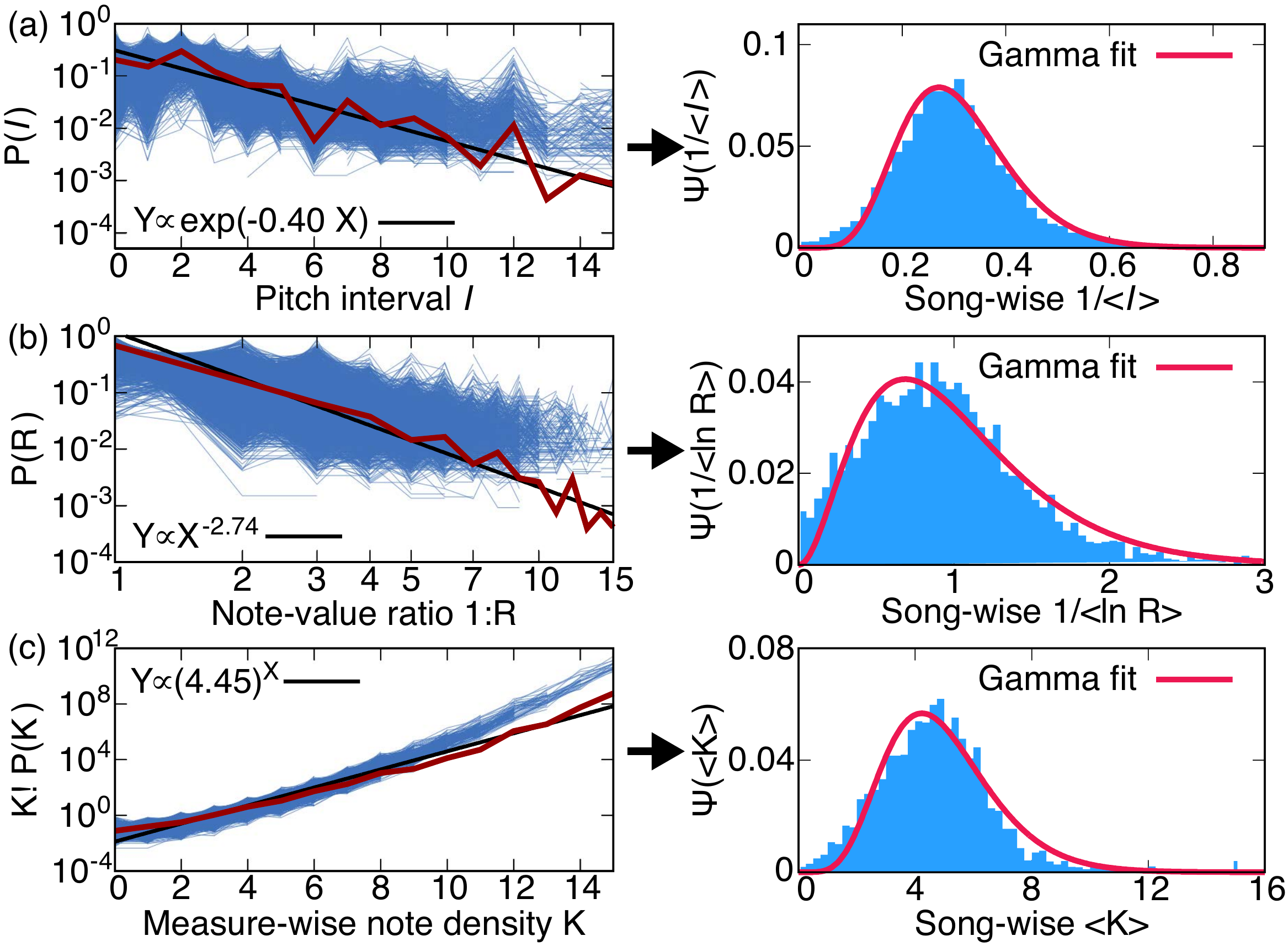}
\caption{Distributions of global musical statistics in the Wikifonia data. (a) The left panel shows the distributions of pitch intervals within songs (fine line segments), that within the whole dataset (bold line segments), and a fitted exponential distribution (straight line); the right panel shows the distribution of song-wise statistics (reciprocals of the means) of pitch intervals and a fitted gamma distribution. (b) Similar results for note value ratios of the form $1$:$R$, where the global distribution is fitted by a Pareto distribution. (c) Similar results for measure-wise note densities, where the global distribution is fitted by a Poisson distribution.}
\label{fig:GlobalStatisticsWikifonia}
\end{figure}
\begin{figure}
\centering
\includegraphics[width=0.95\columnwidth]{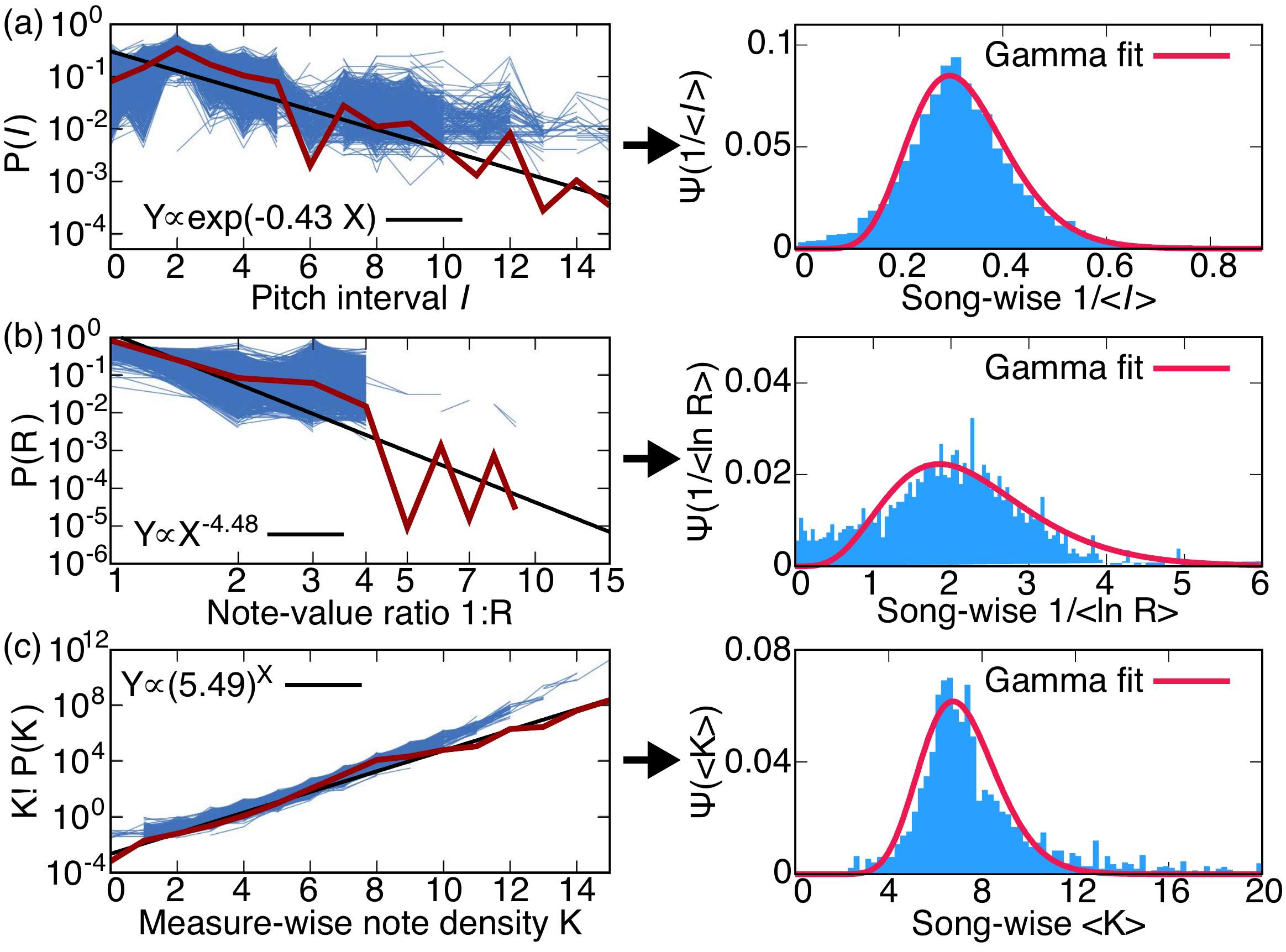}
\caption{Distributions of global musical statistics in the Irish song data. See the caption to Fig.~\ref{fig:GlobalStatisticsWikifonia}.}
\label{fig:GlobalStatisticsIrish}
\end{figure}
The results for the classical music, Wikifonia, and Irish song datasets are shown in Figs.~\ref{fig:GlobalStatisticsClassical}, \ref{fig:GlobalStatisticsWikifonia}, and \ref{fig:GlobalStatisticsIrish}, respectively.
The pitch interval distributions for the classical music data in Fig.~\ref{fig:GlobalStatisticsClassical} significantly differ from the other cases because this dataset contains musical pieces by various instrumentations (including ensemble music) whereas the other three datasets contain (mostly vocal) melodies.
Despite this difference, the tendencies that the contour is approximately exponential distributed and the within-population distribution is approximately gamma distribution are same as the other cases.
The results for the Wikifonia data in Fig.~\ref{fig:GlobalStatisticsWikifonia} are quantitatively similar to the results for the J-pop data in Fig.~\ref{fig:GlobalStatisticsJpop}, indicating that the general tendencies are similar for popular music developed in different regions.
The within-song distributions of the note-value ratios for the Irish music data in Fig.~\ref{fig:GlobalStatisticsIrish}(b) clearly differ from the other cases, reflecting the characteristic of the Irish music that note value ratios $1$:$R$ with $R$ greater than 4 are much rarer compared to the other cases.
The within-population distribution of the shape parameter has a noticeable excess in the small regime in comparison to the fitted gamma distribution.
As small shape parameters correspond to large values of $\langle{\rm ln}\,R\rangle$, the fact that data samples are sparse for large $R$ may have caused unreliable estimation in this regime.
While there are some quantitative and qualitative differences, the feature that the conjugate gamma distributions appear in the within-corpus distributions is found universally for all the analyzed datasets from different musical genres.

We can estimate the values of $D(\bar{\theta})/V(\theta)$ for the global statistics, similarly as for the frequency statistics.
The values estimated from the distributions of (pitch intervals, note value ratios) were ($2.8$, N/A) (classical), ($7.0,2.8$) (Wikifonia), ($10.9,3.1$) (J-pop), and ($7.2,4.8$) (Irish) (measure-wise note density were excluded from the analysis since the units of data samples are different).
These values are significantly smaller than the corresponding values for the frequency statistics, suggesting that there is even larger statistical dependence on the samples (possibly due to global repetitive structure within songs) for these statistics.

%%%%%%%%%%%%%%%%%%%%%%%%%%%%%%%%%%%%%%%%%%%%%%
\section{Examples of exponential family distributions with quadratic estimation variances}
\label{app:ExamplesOfQVDistributions}
%%%%%%%%%%%%%%%%%%%%%%%%%%%%%%%%%%%%%%%%%%%%%%

We here calculate the estimation variances for the well-known distributions listed in Table \ref{tab:Examples} in the expectation value parameterization and confirm that they are constant, linear, or quadratic functions.
We also identify the corresponding conjugate distributions.

\paragraph{Bernoulli.}
The Bernoulli distribution is defined as
\begin{align}
{\rm Ber}(x;p)&=p^x(1-p)^{1-x}
\notag\\
&={\rm exp}\big[x\big\{{\rm ln}\,p-{\rm ln}\,(1-p)\big\}+{\rm ln}\,(1-p)\big],
\label{eq:Bernoulli}
\end{align}
where $x$ takes values in $\{0,1\}$ and $p$ is the probability to get 1.
The distribution has the form of Eq.~(\ref{eq:ExpFamily}) with 
\begin{equation*}
F(x)=x,\quad B(p)={\rm ln}\,p-{\rm ln}\,(1-p),\quad A(p)=-{\rm ln}\,(1-p).
\end{equation*}
Since the mean and variance of the static $F=x$ are $\langle x\rangle = p$ and $V(x)=p(1-p)$, respectively, $p$ is the expectation value parameter ($\theta=p$) and the estimation variance is $D(\theta)=\theta(1-\theta)$.
The conjugate distribution is calculated as
\begin{equation*}
\tilde{\phi}(\theta;\chi,\nu)\propto {\rm exp}\big[\chi B(\theta)-\nu A(\theta)\big]\propto \theta^\chi(1-\theta)^{\nu-\chi},
\end{equation*}
and therefore
\begin{align}
\tilde{\phi}(\theta;\chi,\nu)&=\frac{1}{B(\chi+1,\nu-\chi+1)}\theta^\chi(1-\theta)^{\nu-\chi}
\notag\\
&={\rm Beta}(\theta;\chi+1,\nu-\chi+1).
\end{align}

\paragraph{Poisson.}
The Poisson distribution is defined as
\begin{equation}
{\rm Pois}(x;\lambda)=\frac{\lambda^xe^{-\lambda}}{x!}={\rm exp}(x\,{\rm ln}\,\lambda-\lambda-{\rm ln}\,x!),
\label{eq:Poisson}
\end{equation}
where $x$ takes values in $\mathbb{N}$ and $\lambda$ is the rate parameter.
The distribution has the form of Eq.~(\ref{eq:ExpFamily}) with 
\begin{equation*}
F(x)=x,\quad B(\lambda)={\rm ln}\,\lambda,\quad A(\lambda)=\lambda.
\end{equation*}
Since the mean and variance of the static $F=x$ are $\langle x\rangle = \lambda$ and $V(x)=\lambda$, respectively, $\lambda$ is the expectation value parameter ($\theta=\lambda$) and the estimation variance is $D(\theta)=\theta$.
The conjugate distribution is calculated as
\begin{equation*}
\tilde{\phi}(\theta;\chi,\nu)\propto {\rm exp}\big[\chi B(\theta)-\nu A(\theta)\big]\propto \theta^\chi e^{-\nu\theta},
\end{equation*}
and therefore
\begin{equation}
\tilde{\phi}(\theta;\chi,\nu)=\frac{\nu^{\chi+1}}{\Gamma(\chi+1)}\theta^\chi e^{-\nu\theta}={\rm Gam}(\theta;\chi+1,\nu).
\end{equation}

\paragraph{Gaussian parameterized by mean.}
The Gaussian distribution is defined as
\begin{align}
{\rm Gauss}(x;\mu,\Sigma)&=\frac{1}{\sqrt{2\pi\Sigma}}e^{-\frac{(x-\mu)^2}{2\Sigma}}
\notag\\
&={\rm exp}\bigg[\frac{\mu}{\Sigma}x-\frac{x^2}{2\Sigma}-\frac{\mu^2}{2\Sigma}-\frac{1}{2}\,{\rm ln}(2\pi\Sigma)\bigg],
\label{eq:Gaussian}
\end{align}
where $x$ takes values in $\mathbb{R}$, and $\mu$ and $\Sigma$ are the mean and variance ($\langle x\rangle=\mu$, $\langle (x-\mu)^2\rangle=\Sigma$).
Focusing on the statistic $x$ and treating $\Sigma$ as a known parameter, the distribution has the form of Eq.~(\ref{eq:ExpFamily}) with 
\begin{equation*}
F(x)=x,\quad B(\mu)=\frac{\mu}{\Sigma},\quad A(\mu)=\frac{\mu^2}{2\Sigma}.
\end{equation*}
Since the mean and variance of the static $F=x$ are $\langle x\rangle = \mu$ and $V(x)=\Sigma$, respectively, $\mu$ is the expectation value parameter ($\theta=\mu$) and the estimation variance is $D(\theta)=\Sigma$.
The conjugate distribution is calculated as
\begin{equation*}
\tilde{\phi}(\theta;\chi,\nu)\propto {\rm exp}\big[\chi B(\theta)-\nu A(\theta)\big]\propto {\rm exp}\bigg[\chi\frac{\theta}{\Sigma}-\nu\frac{\theta^2}{2\Sigma}\bigg],
\end{equation*}
and therefore
\begin{align}
\tilde{\phi}(\theta;\chi,\nu)&=\sqrt{\frac{\nu}{2\pi\Sigma}}\,{\rm exp}\bigg[-\frac{\nu}{2\Sigma}\bigg(\theta-\frac{\chi}{\nu}\bigg)^2\bigg]
\notag\\
&={\rm Gauss}\bigg(\theta;\frac{\chi}{\nu},\frac{\Sigma}{\nu}\bigg).
\end{align}

\paragraph{Gaussian parameterized by variance.}
Focusing on the statistic $(x-\mu)^2$ and treating $\mu$ as a known parameter, the Gaussian distribution in Eq.~(\ref{eq:Gaussian}) has the form of Eq.~(\ref{eq:ExpFamily}) with 
\begin{equation*}
F(x)=(x-\mu)^2,\quad B(\Sigma)=-\frac{1}{2\Sigma},\quad A(\Sigma)=\frac{1}{2}\,{\rm ln}(2\pi\Sigma).
\end{equation*}
Since the mean and variance of the static $F(x)$ are $\langle F\rangle = \Sigma$ and $V(F)=2\Sigma^2$, respectively, $\Sigma$ is the expectation value parameter ($\theta=\Sigma$) and the estimation variance is $D(\theta)=2\theta^2$.
The conjugate distribution is calculated as
\begin{equation*}
\tilde{\phi}(\theta;\chi,\nu)\propto {\rm exp}\big[\chi B(\theta)-\nu A(\theta)\big]
\propto \theta^{-\nu/2}e^{-\chi/(2\theta)}
\end{equation*}
and therefore
\begin{equation}
\tilde{\phi}(\theta;\chi,\nu)=\frac{(\frac{\chi}{2})^{\frac{\nu}{2}-1}}{\Gamma(\frac{\nu}{2}-1)}\frac{1}{\theta^{\frac{\nu}{2}}}e^{-\frac{\chi}{2\theta}}
={\rm InvGam}\bigg(\theta;\frac{\nu}{2}-1,\frac{\chi}{2}\bigg).
\end{equation}

\paragraph{Gamma parameterized by mean.}
The Gamma distribution is defined as
\begin{equation}
{\rm Gam}(x;\alpha,\beta)=\frac{\beta^\alpha}{\Gamma(\alpha)}x^{\alpha-1}e^{-\beta x},
\label{eq:Gamma}
\end{equation}
where $x$ takes values in $[0,\infty)$, and $\alpha$ and $\beta$ denote the shape and rate parameters.
Focusing on the statistic $x$ and treating $\alpha$ as a known parameter, the distribution has the form of Eq.~(\ref{eq:ExpFamily}) with 
\begin{equation*}
F(x)=x,\quad B(\beta)=-\beta,\quad A(\beta)=-\alpha\,{\rm ln}\,\beta.
\end{equation*}
Since the mean and variance of the static $F=x$ are $\langle x\rangle =\alpha/\beta$ and  $V(x)=\alpha/\beta^2$, respectively, $\theta\equiv\alpha/\beta$ is the expectation value parameter and the estimation variance is given as $D(\theta)=\theta^2/\alpha$.
The conjugate distribution is calculated as
\begin{equation*}
\tilde{\phi}(\theta;\chi,\nu)\propto {\rm exp}\big[\chi B(\theta)-\nu A(\theta)\big]\propto {\rm exp}\bigg[-\frac{\alpha\chi}{\theta}-\alpha\nu\,{\rm ln}\,\theta\bigg]
\end{equation*}
and therefore
\begin{equation}
\tilde{\phi}(\theta;\chi,\nu)=
\frac{(\chi\alpha)^{\nu\alpha-1}}{\Gamma(\nu\alpha-1)}\frac{e^{-\chi\alpha/\theta}}{\theta^{\nu\alpha}}={\rm InvGam}(\theta;\nu\alpha-1,\chi\alpha).
\end{equation}
With the relation $\beta=\alpha/\theta$, this means that $\beta$ is gamma distributed:
\begin{equation}
\tilde{\phi}(\beta;\chi,\nu)={\rm Gam}(\beta;\nu\alpha-1,\chi).
\end{equation}

The exponential distribution is a special case of gamma distribution and the Pareto distribution can be obtained from the exponential distribution by a logarithmic conversion.
Therefore, the above calculation shows that the conjugate distributions for the statistical parameters analyzed in Figs.~\ref{fig:GlobalStatisticsJpop}(a) and (c) are gamma distributions.

%%%%%%%%%%%%%%%%%%%%%%%%%%%%%%%%%%%%%%%%%%%%%%
\section{Analysis of the solution of the FP equation (\ref{eq:FP})}
\label{app:AnalysisOfEquilibriumDistribution}
%%%%%%%%%%%%%%%%%%%%%%%%%%%%%%%%%%%%%%%%%%%%%%

We first calculate the mean $\langle\theta\rangle_*\equiv\int d\theta\,\theta\Psi_*(\theta)$ for the zero-current equilibrium solution $\Psi_*$ to Eq.~(\ref{eq:FP}).
Integrating both sides of Eq.~(\ref{eq:ZeroCurrentEquation}) gives
\begin{equation*}
2Lu(\langle\theta\rangle_*-\zeta)=-[D\Psi_*]^{\theta_+}_{\theta_-},
\end{equation*}
where $(\theta_-,\theta_+)$ represents the domain of $\theta$.
If the surface term on the RHS vanishes at both ends of the domain of $\theta$, as it regularly does, we have
\begin{equation*}
\langle\theta\rangle_*=\zeta.
%\label{eq:EquilibriumMean}
\end{equation*}
The physical meaning of this relation is clear: after a long time the directional effect of the oblique transmission (linear pressure) term $M(\theta)$ dominates (the effect of the diffusion term $K(\theta)$ is directionless).

Next, we suppose that data production model $\phi$ has a quadratic estimation variance as in Eq.~(\ref{eq:QuadraticCondition}).
Using Eq.~(\ref{eq:ZeroCurrentEquation}), we have
\begin{align}
&2Lu\langle(\theta-\zeta)^2\rangle_*=2Lu\int d\theta\,(\theta-\zeta)^2\Psi_*
\notag\\
&=-\int d\theta\,(\theta-\zeta)\frac{d}{d\theta}(D\Psi_*)
\notag\\
&=-\big[(\theta-\zeta)D\Psi_*\big]^{\theta_+}_{\theta_-}+\int d\theta\,D\Psi_*.
\label{eq:CalculationEquilibriumVariance}
\end{align}
The last integral can be transformed as
\begin{equation*}
\int d\theta\,D\Psi_*=d_0+d_1\langle\theta\rangle_*+d_2\langle\theta^2\rangle_*=D(\zeta)+d_2\langle(\theta-\zeta)^2\rangle_*.
\end{equation*}
Thus, if the surface term in Eq.~(\ref{eq:CalculationEquilibriumVariance}) vanishes at both ends of the domain of $\theta$, as it regularly does, we have
\begin{equation}
\langle(\theta-\zeta)^2\rangle_*=\frac{D(\zeta)}{2Lu-d_2},
\label{eq:EquilibriumVariance}
\end{equation}
which is same as Eq.~(\ref{eq:VarianceAtEquilibrium}).
This result can be explicitly checked for the example cases presented in the previous section by using the equilibrium solution in Eq.~(\ref{eq:ExplicitEquiSol}) and the known formulas for the variances of the conjugate distributions.

Given the stationary solution $\Psi_*(\theta)$, the time-dependent FP equation (\ref{eq:FP}) can in general be solved by the eigenfunction method.
Here, we use this method to calculate the convergence time to the equilibrium.
We define the coefficient function $Q(\theta,t)$ by $\Psi(\theta,t)=Q(\theta,t)\Psi_*(\theta)$ and substitute it into Eq.~(\ref{eq:FP}) to obtain
\begin{equation*}
\frac{\partial Q(\theta,t)}{\partial t}=\bigg[M(\theta)\frac{\partial}{\partial\theta}+\frac{K(\theta)}{2}\frac{\partial^2}{\partial\theta^2}\bigg]Q(\theta,t).
%\label{eq:FPEForCoefficient}
\end{equation*}
Thus, the equation for an eigenfunction $Q(\theta,t)=e^{-\lambda t}Q_\lambda(\theta)$ is given as
\begin{equation*}
\bigg[\frac{K(\theta)}{2}\frac{d^2}{d\theta^2}+M(\theta)\frac{d}{d\theta}+\lambda\bigg]Q_\lambda(\theta)=0.
%\label{eq:EigenfuncEq}
\end{equation*}
Using the method of series expansion $Q_\lambda=\sum_{n=0}^\infty a_n\theta^n$ and substituting $M=u(\zeta-\theta)$ and $K=(d_0+d_1\theta+d_2\theta^2)/L$, we obtain the following iterative equation for the coefficients:
\begin{align}
&d_0(n+1)(n+2)a_{n+2}+\big\{d_1n(n+1)+2L(n+1)u\zeta\big\}a_{n+1}
\notag\\
&+\big\{d_2n(n-1)+2L(\lambda-nu)\big\}a_n=0.
\label{eq:CoeffEq}
\end{align}
For all the distributions listed in Table \ref{tab:Examples}, one can check that the series $Q_\lambda$ must have a finite order to be a normalizable solution.
This leads to the condition that the coefficient of $a_n$ in Eq.~(\ref{eq:CoeffEq}) must vanish for some $n$, and we obtain discrete eigenvalues
\begin{equation}
\lambda_m=mu-\frac{m(m-1)}{2L}d_2\quad (m=0,1,2,\ldots).
\end{equation}
Since the lowest modes are $\lambda_0=0$, $\lambda_1=u$, $\cdots$, the lowest nonzero mode decays as $\propto e^{-ut}$.
That is, the relaxation time (convergence time to the equilibrium) is of order $u^{-1}$, reproducing the physically expected result as explained in the main text.

%%%%%%%%%%%%%%%%%%%%%%%%%%%%%%%%%%%%%%%%%%%%%%
\section{Numerical analysis of the solution of the FP equation (\ref{eq:FP})}
\label{app:NumericalAnalysisOfEquilibriumDistribution}
%%%%%%%%%%%%%%%%%%%%%%%%%%%%%%%%%%%%%%%%%%%%%%

%
\begin{figure}
\centering
\includegraphics[width=0.95\columnwidth]{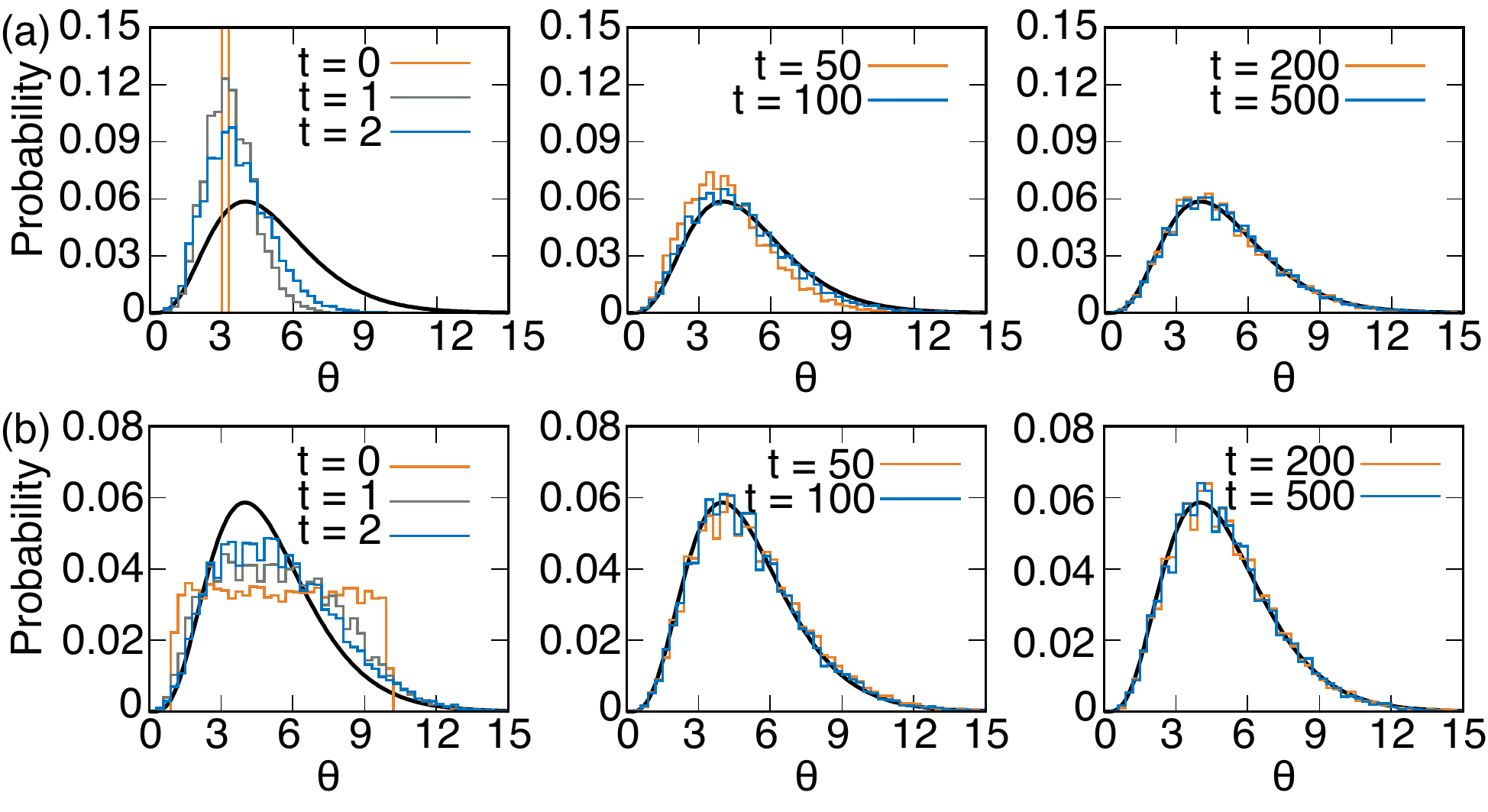}
\caption{Simulation results for the SLG system with Poisson distributed production models. Parameters were set as $N=20000$, $L=25$, $u=1/50=0.02$, and $\zeta=5$. The fitted distribution is ${\rm Gam}(\theta;5,1)$. (a) Initial distribution concentrated at $\theta=3$. (b) Initial distribution uniformly distributed in the range $[1,10]$.}
\label{fig_Sim_PoissonSLGS}
\end{figure}
\begin{figure}
\centering
\includegraphics[width=0.95\columnwidth]{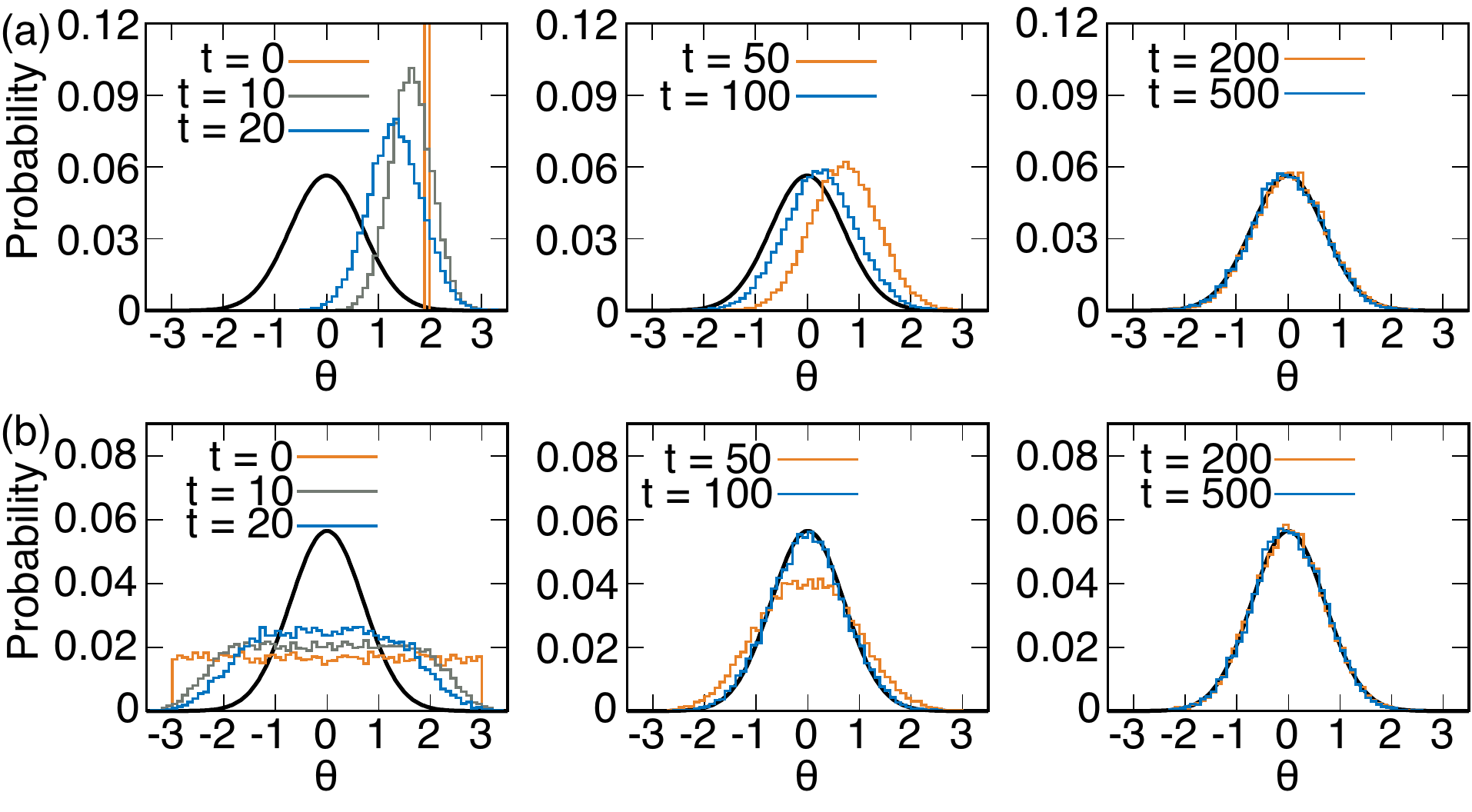}
\caption{Simulation results for the SLG system with Gaussian distributed production models and mean value traits with a fixed variance $\Sigma=1$. Parameters were set as $N=20000$, $L=50$, $u=1/50=0.02$, and $\zeta=0$. The fitted distribution is ${\rm Gauss}(\theta;0,0.5)$. (a) Initial distribution concentrated at $\theta=2$. (b) Initial distribution uniformly distributed in the range $[-3,3]$.}
\label{fig_Sim_GaussianMeanSLGS}
\end{figure}

To confirm the analytical results by numerical calculation, we conducted simulation experiments for statistical learn-generate (SLG) systems whose data production models are one of those listed in Table \ref{tab:Examples}.
For each SLG system, we iterate the learning and generation processes with oblique transmission (with constant $\zeta$) and observe that the within-population distribution of the cultural traits (statistical parameters) converge to the conjugate distribution with parameters given in Eq.~(\ref{eq:ExplicitEquiSol}).
We also confirm that the relaxation time is of order $u^{-1}$.

For simulation, we tested two extreme cases for the initial within-population distribution, a delta function and a uniform distribution with a certain value range.
The data generation process was simulated by using the standard methods for sampling from the probability distributions implemented in the \verb$C++$ standard library under the header \verb$<random>$ (the 64-bit Mersenne Twister was used for random number generation).
To get samples from Poisson distributions, we used Knuth's algorithm (D.~E.~Knuth, ``The Art of Computer Programming 2. Seminumerical Algorithms (3rd ed.),'' Addison Wesley, 1997).

\begin{figure}
\centering
\includegraphics[width=0.95\columnwidth]{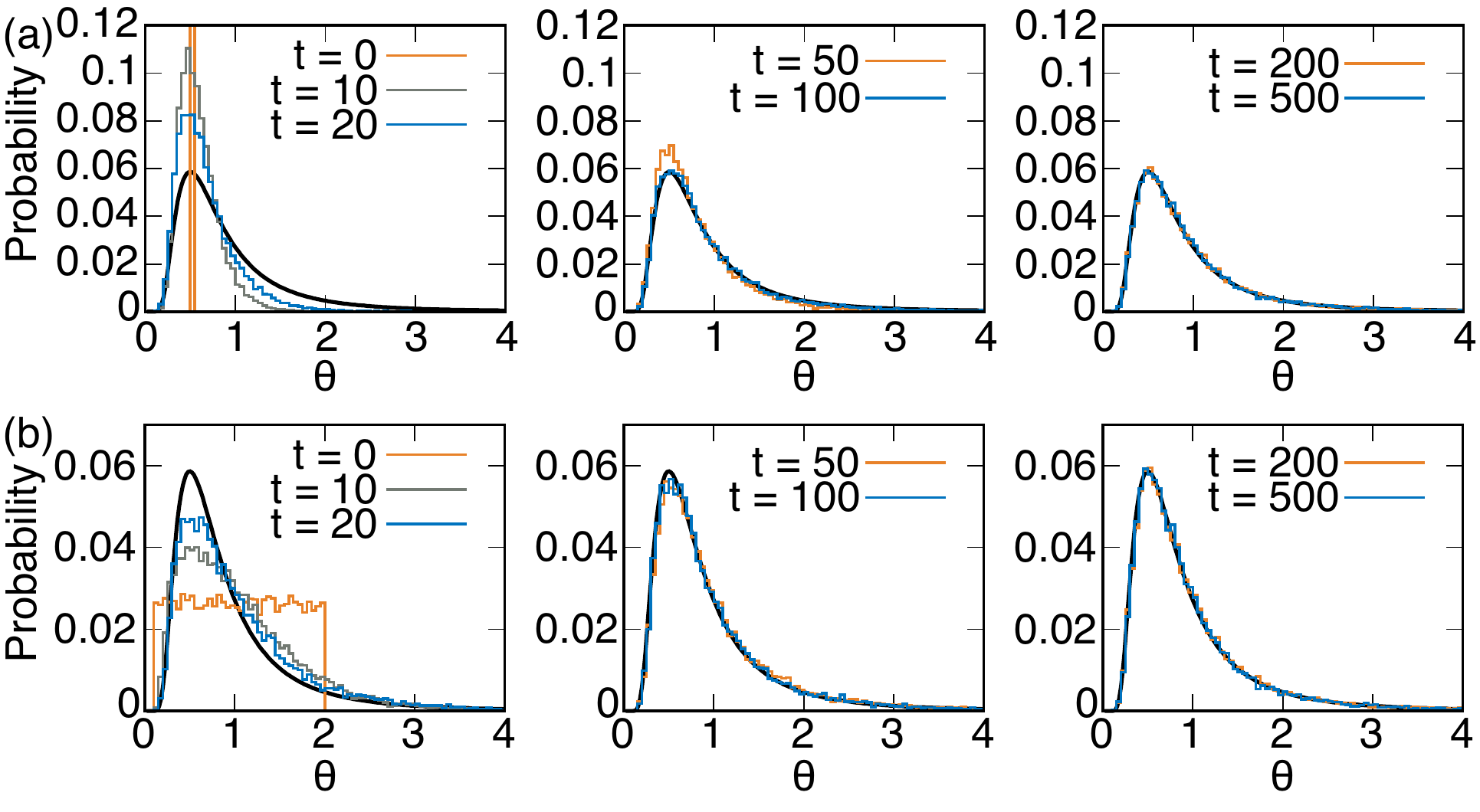}
\caption{Simulation results for the SLG system with Gaussian distributed production models and variance value traits with a fixed mean $\mu=0$. Parameters were set as $N=20000$, $L=100$, $u=1/50=0.02$, and $\zeta=1$. The fitted distribution is ${\rm InvGam}(\theta;3,2)$. (a) Initial distribution concentrated at $\theta=0.5$. (b) Initial distribution uniformly distributed in the range $[0.1,2]$.}
\label{fig_Sim_GaussianVarianceSLGS}
\end{figure}
\begin{figure}
\centering
\includegraphics[width=0.95\columnwidth]{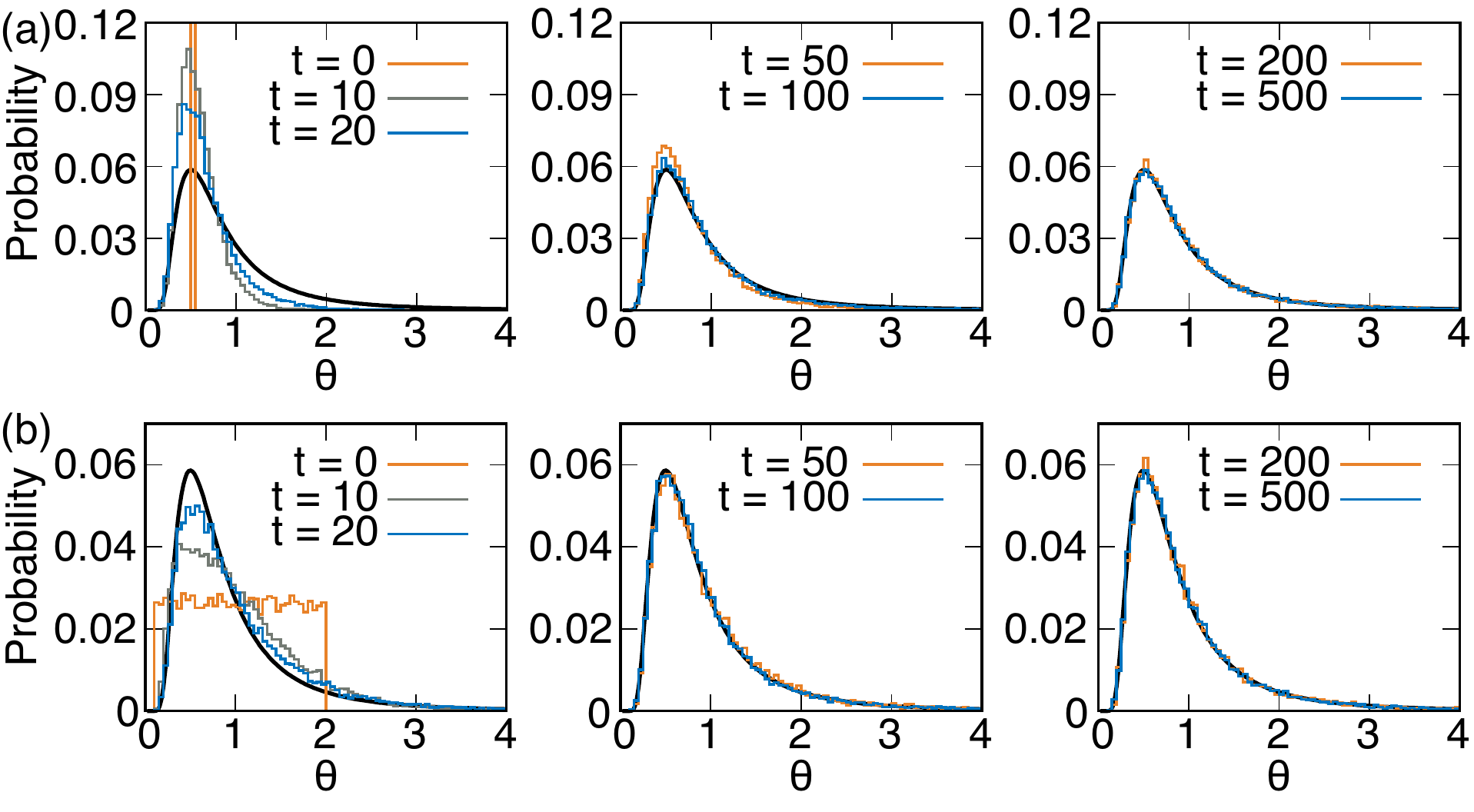}
\caption{Simulation results for the SLG system with gamma distributed production models and mean value traits $\theta$ with a fixed shape parameter $\alpha=0$. The rate parameter is $\beta=1/\theta$. Parameters were set as $N=20000$, $L=50$, $u=1/50=0.02$, and $\zeta=1$. The fitted distribution is ${\rm InvGam}(\theta;3,2)$. (a) Initial distribution concentrated at $\theta=0.5$. (b) Initial distribution uniformly distributed in the range $[0.1,2]$.}
\label{fig_Sim_GammaMeanSLGS}
\end{figure}

The results are shown in Fig.~\ref{fig_Sim_BernoulliSLGS} (in the main text) and Figs.~\ref{fig_Sim_PoissonSLGS} to \ref{fig_Sim_GammaMeanSLGS}.
In all cases, we see that the within-population distribution converges to the theoretically predicted conjugate distribution regardless of the initial distribution.
It can be also confirmed that the convergence is attained at time $t\sim100=2u^{-1}$.

%%%%%%%%%%%%%%%%%%%%%%%%%%%%%%%%%%%%%%%%%%%%%%
\section{Equilibrium solution of the FP equation in the multi-parameter case}
\label{app:FPEqMultiParameters}
%%%%%%%%%%%%%%%%%%%%%%%%%%%%%%%%%%%%%%%%%%%%%%

We here present the detailed derivation of the result in Sec.~\ref{sec:FPEqMultiParameters}.
We use the same model and notation described there.
Differentiating Eq.~(\ref{eq:MultiParamEVParameterization}), we have $G_{ij}=\partial_i\partial_jA-\theta_k\partial_i\partial_jB_k$.
From $\langle\partial_i\partial_j\phi\rangle=0$, we obtain,
\begin{equation*}
\partial_i\partial_j A-\langle F_k\rangle\partial_i\partial_j B_k=\langle (F_kG_{ik}-\partial_iA)(F_\ell G_{j\ell}-\partial_jA)\rangle.
\end{equation*}
Multiplying $G^{-1}_{mi}G^{-1}_{nj}$ on both sides and summing over $i$ and $j$, we obtain Eq.(\ref{eq:EstimationCovariance}) as
\begin{equation*}
D_{mn}\equiv\langle (F_m-\theta_m)(F_n-\theta_n)\rangle
=G^{-1}_{mi}G^{-1}_{nj}G_{ij}=G^{-1}_{mn}.
\end{equation*}

Next, we solve the zero-current equation (\ref{eq:ZeroCurrentEqMultiParam}).
We assume the following conjugate distribution: $\Psi\propto{\rm exp}[\chi_kB_k(\theta)-\nu A(\theta)+C(\theta)]$.
Since
\begin{equation*}
\partial_j\Psi=(\chi_k\partial_jB_k-\nu\partial_jA+\partial_jC)\Psi
=\big[G_{jk}(\chi_k-\nu\theta_k)+\partial_jC\big]\Psi,
\end{equation*}
Eq.~(\ref{eq:ZeroCurrentEqMultiParam}) is equivalent to
\begin{equation*}
\big\{2Lu\zeta_i-\chi_i+(\nu-2Lu)\theta_i-\partial_jD_{ij}-D_{ij}\partial_jC\big\}\Psi=0.
\end{equation*}
Thus, the conjugate distribution is a solution if the following equations are satisfied:
\begin{equation*}
\nu=2Lu,\quad \chi_i=2Lu\zeta_i,\quad \partial_jD_{ij}+D_{ij}\partial_jC=0.
\end{equation*}
The last equation is equivalent to
\begin{equation*}
\partial_k C=-G_{ki}\partial_jD_{ij}.
\end{equation*}
Since $D=G^{-1}$, $\partial_j D=-D(\partial_jG)D$ or $\partial_jD_{k\ell}=-G^{-1}_{km}G^{-1}_{n\ell}\partial_jG_{mn}$.
The RHS can be transformed as
\begin{align*}
-G_{ki}\partial_jD_{ij}&=G^{-1}_{ij}\partial_jG_{ki}=G^{-1}_{ij}\partial_j\partial_kB_i
\notag\\
&=G^{-1}_{ij}\partial_k\partial_jB_i=G^{-1}_{ij}\partial_kG_{ji},
\end{align*}
and the equation for $C$ is
\begin{equation*}
\partial_kC=G^{-1}_{ij}\partial_kG_{ji}.
\end{equation*}
Using Jacobi's formula this can be solved as $C={\rm ln}\,{\rm det}\,G$.
Therefore, the equilibrium solution is given as in Eq.~(\ref{eq:EquivSolutionMultiParam}).

\end{document}